\documentclass[aps,superscriptaddress,preprint,pra,showpacs]{revtex4}
\usepackage{amsmath}
\usepackage{graphicx}
\usepackage{graphicx,color}	
\usepackage{amsfonts}
\usepackage{amssymb}
\newcommand*{\Eq}[1]{Eq.~(\ref{#1})}
\newcommand*{\ket}[1]{|#1\rangle}
\newcommand*{\bra}[1]{\langle#1|}
\newcommand*{\bracket}[1]{\langle#1\rangle}
\newcommand*{\ketbra}[1]{\ket{#1}\bra{#1}}
\newcommand*{\bA}{\mathbf{A}}
\newcommand*{\bS}{\mathbf{S}}
\newcommand*{\cE}{\mathcal{E}}

\newcommand*{\cH}{\mathcal{H}}
\newcommand*{\Tr}{\mathrm{Tr}}
\newcommand*{\beqa}{\begin{eqnarray}}
\newcommand*{\eeqa}{\end{eqnarray}}
\newcommand*{\beqas}{\begin{eqnarray*}}
\newcommand*{\eeqas}{\end{eqnarray*}}
\newcommand*{\rank}{\mathrm{rank}}

\begin{document}

\title{Conservation-Law-Induced Quantum Limits for
Physical Realizations of the Quantum NOT Gate}

\author{Tokishiro Karasawa}
\email[]{jidai@nii.ac.jp}
\affiliation{National Institute of Informatics, 
Chiyoda-ku, Tokyo,  101-8430, Japan}

\author{Masanao Ozawa}
\email[]{ozawa@math.is.tohoku.ac.jp}
\affiliation{Graduate School of Information Sciences,
T\^{o}hoku University, Aoba-ku, Sendai,  980-8579, Japan}

\begin{abstract}
In recent investigations, it has been found that conservation laws  
generally lead to precision limits
on quantum computing. 
Lower bounds of the error probability have been  
obtained for various logic
operations from the commutation relation between the noise operator  
and the conserved quantity
or from the recently developed universal uncertainty principle for  
the noise-disturbance trade-off
in general measurements. 
However, the problem of obtaining the precision limit to realizing
the quantum NOT gate has eluded
a solution from these approaches. 
Here, we  develop a new
method for this problem based on analyzing the trace distance between  
the output state from the
realization under consideration and the one from the ideal gate.
Using the mathematical
apparatus of orthogonal polynomials, we obtain a general lower bound  
on the error probability for
the realization of the quantum NOT gate in terms of the number of  
qubits in the control system
under the conservation of the total angular momentum of the computational qubit
plus the the control system along the direction used to encode the computational basis.
The lower bound turns out to be more stringent than one might expect from  
previous results. 
The new method is expected to lead to more accurate estimates for  
physical realizations of
various types of quantum computations under conservation laws, and to  
contribute to related problems
such as the accuracy of programmable quantum processors.
\end{abstract}


\maketitle

\section{Introduction}
Recently, there have been extensive research efforts to explore whether fundamental 
physical laws put any constraints on realizing scalable quantum computing.
Soon after the discovery of Shor's algorithm \cite{Sho94}, it was pointed out by several
physicists \cite{Unr95, PSE96, HR96} that the decoherence, the exponential decay of
coherence in time, caused by the  coupling between a quantum computer and the environment
would cancel out the computational advantage of quantum computers.  
To overcome this difficulty, quantum error-correction was proposed \cite{Sho95b, Ste96},
and the subsequent development has established the so-called threshold theorem:
if the error caused by the decoherence in individual quantum gates is 
below a certain constant threshold,
it is possible in principle to efficiently perform an arbitrary scale 
of fault-tolerant quantum computation with
error-correction \cite{NC00}.
Thus, the error-correction reduces, in principle,
the scalability problem to the accuracy problem 
requiring individual quantum logic gates
to clear the error threshold, though being still 
quite demanding.  

In general, decoherence in quantum computer components
can be classified into two classes: (i) static decoherence,
arising from the interaction between
computational qubits, typically in the memory, and the
environment, and (ii) dynamical decoherence,
arising from the interaction between computational
qubits, typically in the register, and the control system
of gate operations \cite{03QLM}.
The static decoherence may be overcome by
developing materials with long decoherence time.
On the other hand, the dynamical decoherence
poses a dilemma between controllability and decoherence;
the control needs coupling, whereas the coupling causes
decoherence. 
Thus, even if the interaction with the environment is 
completely suppressed, the error caused 
by the dynamical decoherence still remains. 
Clearly, if the control system is described classically, there is no 
decoherence. However, this never happens in reality with finite
resources.

Barnes, Warren \cite{BW99}, Gea-Banacloche \cite{Ban02},  
van Enk, and Kimble \cite{EK02} 
have been focused on the atom-field interaction
between atom qubits and control electromagnetic fields,
and shown that, when the control field is in a coherent state, 
the gate error scales as the inverse of the average photon number.
In contrast to those model-dependent approaches, one of the authors 
\cite{02CQC}  explored the physical constraint on the error caused by 
dynamical decoherence generally imposed by conservation laws 
and obtained accuracy limits by
quantitatively generalizing the so-called Winger-Araki-Yanase theorem
\cite{Wig52,AY60}:
observables which do not commute with bounded additive
conserved quantities have no precise and non-disturbing measurements.
It is natural to assume that conservation laws are satisfied by the interaction
between the qubit and the external control system.
If the control system were to be completely described as a classical system, 
the conservation law would not cause any conflict in realizing a unitary operation
on the computational qubit, since 
the classical interaction causes no decoherence and yet 
conserves the (infinite) total quantum number.  
However, in reality, the interaction may cause
decoherence and the time evolution  operator on the composite 
system is limited to one commuting with the conserved quantity.   
Under these conditions,  the accuracy of the realized gate operation
generally depends on the kind of gate being considered.  
It has been shown that the SWAP gate can be realized in principle without error \cite{02CQC}.  
However, the controlled NOT gate and the Hadamard gate have lower bounds 
of the error probability that scales as the inverse of the size of the control system, as follows.

The impossibility of precise and non-disturbing measurements under conservation laws
was generalized to an inequality for
the lower bound of  the sum  of the noise and the disturbance of measuring  process 
under a conservation law \cite{03QLM}.
This inequality leads to a general lower bound for the error probability 
of any realization of the controlled-NOT gate under conservation laws 
\cite{02CQC,03QLM,03CQC(R)}.
For single-spin qubits controlled by the $N$-qubit control system,
the angular momentum conservation law leads to the minimum error probability 
$(4N^{2})^{-1}$ \cite{02CQC}.  Thus, assuming the threshold error probability 
$10^{-4}-10^{-5}$ \cite{NC00}, a two-qubit unitary operator needs to be realized 
by an interaction with more than 100 qubit systems, 
suggesting the usefulness of schemes based on multiple-spin encoded qubits 
such as the universal encoding based on decoherence-free subspaces 
\cite{Lid03,03CQC(R),06QGG}.
In bosonic controls, such as electromagnetic fields in coherent states, 
the minimum error probability  amounts to $(16\bar{n})^{-1}$ \cite{02CQC}, 
where $\bar{n}$ is the average number of photons.
The above result also leads to a conclusion that in any universal set of elementally logic operations
there is at least one logic operation that obeys the error limit with the same scaling as above 
\cite{02CQC}.  

On the other hand, without assuming the non-disturbing condition the lower bound for
the noise in arbitrary measurements under arbitrary conservation laws was derived from the
commutation relation for noise operator and the conserved quantity \cite{02CLU} or 
simply from the universal uncertainty principle \cite{03UPQ};
see Refs.~\cite{03UVR,03HUR,04URN} for the universal uncertainty principle.
This inequality also leads to a general lower bound for the error probability of the 
realization of the Hadamard gate
that amounts to the minimum error probability $(4N^{2})^{-1}$ for any $N$-qubit
control system and $(16\bar{n})^{-1}$ for any electromagnetic control field in
a coherent state with average number of photons $\bar{n}$  \cite{03UPQ}.
Gea-Banacloche and one of the authors \cite{05CQL} compared the 
above result for electromagnetic control fields
with the previous result obtained by Gea-Banacloche \cite{Ban02} for the
Jaynes-Cummings interaction, and  it was concluded that the constraint 
based on the angular momentum conservation law represents an ultimate limit 
closely related to the fluctuations in the quantum field phase.
The use of the Jaynes-Cummings model in the above model-dependent approach
\cite{Ban02,EK02} was questioned by Itano \cite{Ita03} and subsequently 
Silberfarb and Deutsch \cite{SD04} justified the Jaynes-Cummings model
in the limit of small entanglement; see also replies to Itano by van Enk 
and H. J. Kimble \cite{EK03} and by Gea-Banacloche \cite{Ban03}.
The above consistency result between the model-dependent and model-independent 
approaches enforces the validity of the use of the Jaynes-Cummings model
and substantially clarifies the whole situation.

The above methods for deriving conservation-law-induced quantum limits for 
quantum logic operations are also applicable to the Toffoli gate and the Fredkin 
gate to obtain similar lower bounds.  
However, the problem of obtaining the precision limit to realizing
the quantum NOT gate has eluded a solution from these approaches,
and hence the problem has been open as to how
the minimum error for that gate scales with the size of the control system.
In this paper, in order to solve this problem we devise a new method 
of deriving the precision limit, 
and show that there exists a non-zero lower bound, 
which indeed scales as the inverse size of the control system,
of the error probability  for the quantum NOT gate. 

Our formulation has various common features with the formulation 
of programmable quantum processors \cite{NC97,VC00,HZB06}, 
in which a set of unitary operators is to be realized by
selecting a unitary operator on the composite system,
the system plus the ancilla, and by selecting a set
of ancilla states, whereas in our problem a single unitary operator is to be
realized under a conservation law by selecting a unitary operator on the
composite system satisfying the conservation laws and by selecting a single
ancilla state.  
In previous investigations the accuracy of programmable 
quantum computing has been measured by the so-called process fidelity,
a fidelity based distance measure between two operations, 
whereas here we investigate in the completely bounded (CB) distance 
or gate trace distance, a trace-distance based measure.  
Thus, our method is expected to contribute to the problem of 
programmable quantum processors and related subjects
\cite{DP05,DP05b,DP05c} in future investigations.

The paper is organized as follows.
Sec.~\ref{def-error-pro} gives basic formulations and 
main results.  We define the error probability in realizing 
the quantum NOT gate based on the CB distance. 
We subsequently show that a pure input
state gives the worst error probability.  
This enables us to assume, without loss of generality, 
that the input state is a pure state. 
In preparation for deriving the lower bound of the error probability,
in Sec.~\ref{Der-of-the-} we generally describe the maximum trace distance 
between the two output states from the realization 
and from the ideal quantum NOT gate.
In Sec.~\ref{lowerbound-conservation}, 
we introduce the conservation law into the discussion. 
By minimizing the error probability
over arbitrary choices of  the evolution operator obeying the conservation law, 
we give a lower bound which depends only on the ancilla input state. 
In Sec.~\ref{accuracy-limit-size}, we optimize the ancilla input state and 
derive a general lower bound expressed as a function of the size (the number of qubits) of the
ancilla.  
Chebyshev polynomials of the second kind, a family of orthogonal polynomials,  
are used to solve this problem. 
To show the tightness of the bound, 
in Sec.~\ref{a-noteworthy}, we consider 
classically complete realizations,
realizations which correctly carry out the quantum NOT operation when
the input state is a computational basis state.
We obtain the attainable lower bound for classically complete realizations.
This result also shows that the general lower bound
can be attained up to constant factor of the ancilla size.
In the final section, we summarize our study and comment 
on the direction of future studies. 

\section{Formulation and main results \label{def-error-pro}}

\subsection{Qubits and conservation laws}

The problem to be considered is formulated as follows.
The main system $\bS$ is a single qubit described 
by a two dimensional Hilbert space $\cH_{\bS}$ with a fixed computational basis
$\{ \ket{0}, \ket{1} \}$.
The Pauli operators $X_{\bS},Y_{\bS}$, and $Z_{\bS}$  
on $\cH_{\bS}$ are defined by
$X_{\bS}=\ket{0}\bra{1} + \ket{1}\bra{0}$, 
$Y_{\bS}=- i \ket{0}\bra{1} + i \ket{1}\bra{0}$, and
$Z_{\bS}=\ket{0}\bra{0} - \ket{1}\bra{1}$.
We refer to $X_{\bS}$ as the quantum NOT gate.

We suppose that the computational basis is represented by the $z$-component of spin, 
and consider the constraint on realizing the quantum NOT gate $X_{\bS}$ 
under the angular momentum conservation law. 
More specifically, we assume that the control system is described 
as an $N$-qubit system $\bA$ also called the ancilla, 
and that the interaction between $\bS$ and $\bA$ preserves the $z$-component of the
angular momentum of the composite system $\bS+\bA$,
and study the unavoidable error probability in realizing the quantum NOT operation.
 
Each qubit ${\bA}_{i}$ for $i=1,2,\cdots,N$ in the ancilla $\bA$ is described 
by a two dimensional Hilbert space
$\cH_{{\bA}_{i}}$.  
Accordingly, the Hilbert space 
$\cH_{\bA}$ of the ancilla $\bA$ is the tensor product 
$
\cH_{\bA}=\otimes^{N}_{i=1} \cH_{{\bA}_{i}}
$,
and 
the Hilbert space $\cH$ of the composite system $\bS+\bA$ is
$
\cH = \cH_{{\bS}} \otimes \cH_{\bA}.
$
The observable $Z_{\bS}$ on $\cH_{{\bS}}$ 
is identified with $ Z_{{\bS}} \otimes 
I_{\bA_{1}} \otimes I_{\bA_{2}} \otimes \cdots 
\otimes I_{\bA_{N}}, \label{Z_S-Z_barS}
$
where $I_{{\bA}_{i}}$ for $i=1,2,\cdots,N$ is the identity operator on 
$\cH_{{\bA}_{i}}$,
respectively. Let $Z_{\bA_{i}}$ be the Pauli Z operator on 
$\cH_{{\bA}_{i}}$,
which is also identified with the corresponding operator on 
$\cH$.
The sum of Pauli Z operators on $\bA$ is denoted by
$$
 {Z}_{\bA}=\sum_{i=1}^{N}{Z}_{{\bA}_{i}},
$$ 
and the corresponding sum of $\bS+\bA$ is denoted by
$$
Z={Z}_{{\bS}} +  {Z}_{\bA}.
$$

Let $U$ be the evolution operator of $\bS+\bA$ during the interaction between 
$\bS$ and $\bA$ to realize the quantum NOT gate on $\bS$.
We assume that $U$ satisfies 
the conservation law 
\begin{eqnarray}
[U,Z] = 0, \label{c.l.}
\end{eqnarray}
where $[U,Z]=UZ-ZU$. 
We shall 
show that the conservation law (\ref{c.l.}) causes unavoidable decoherence in 
realizing $X_{\bS}$ by $U$.

To  obtain the error probability,  
we describe the output state of $\bS$ resulting from
the evolution of $\bS+\bA$. 
Let $\rho_{\bS}$ and $\rho_{\bA}$ be states of $\bS$ and $\bA$, respectively,
so that the input state of $\bS+\bA$ is the product state $\rho_{\bS} \otimes \rho_{\bA}$.
Then the output state $\cE_{U,\rho_{\bA}}(\rho_{\bS})$ 
of $\bS$ is given by
\begin{eqnarray}\label{eq:operation}
\cE_{U,\rho_{\bA}}(\rho_{\bS})=
{\mathrm{Tr}}_{\bA}
\left[ U\left( \rho_{{\bS}} \otimes \rho_{\bA} \right)U^{\dagger}\right],
\label{E_STr_A[Urho^2U]}
\end{eqnarray}
where ${\mathrm{Tr}}_{\bA}\left[ \cdot \right] $ is the partial trace over $\cH_{\bA}$.
On the other hand, for the perfect quantum NOT gate,
the output state $\cE_{X_{\bS}}(\rho_{\bS})$ of $\bS$ would be 
\begin{eqnarray}
\cE_{X_{\bS}}(\rho_{\bS})
&=&X_{\bS}\rho_{\bS}{X_{\bS}}^{\dagger}.
\label{E_XTr_A[Urho^2U]}
\end{eqnarray}

In the following sections we shall show that there exists an unavoidable 
error probability of the output state (\ref{E_STr_A[Urho^2U]}) in
realizing the output state (\ref{E_XTr_A[Urho^2U]}) under the conservation law (\ref{c.l.}). 
The unavoidable error probability for any unitary operator $U$ satisfying 
the conservation law (\ref{c.l.})
will be evaluated to be at least
$$
\frac{1}{2} \Big( 1 -   \cos  \frac{\pi}{  N+2  } \Big)  
$$
for the worst input state $\rho_{\bS}$ of $\bS$ and for the best input state 
$\rho_{\bA}$ of $\bA$, 
and the achievability to this lower bound will be shown asymptotically.
This lower bound is much tighter than the lower bound 
$\frac{1}{16N^2+4}$ anticipated 
from the previous investigations for other gates as to be shown numerically.

\subsection{Error probability and CB distance}

To state our results more precisely, we introduce the following definitions.
Any pair $(U,\rho_{\bA})$ consisting of a unitary operator $U$ on 
$\cH_{\bS}\otimes\cH_{\bA}$ and a state $\rho_{\bA}$ on $\cH_{\bA}$ is called 
a gate implementation or simply an implementation with ancilla $\bA$. 
Every implementation $(U,\rho_{\bA})$ determines the trace-preserving 
completely positive (CP) map
$\cE_{U,\rho_{\bA}}$  of the states of $\bS$
by \Eq{E_STr_A[Urho^2U]} called the gate operation determined by $(U,\rho_{\bA})$;
see Ref.~\cite{NC00} for trace-preserving CP maps in quantum information theory.
An implementation $(U,\rho_{\bA})$ is said to be conservative if it satisfies \Eq{c.l.}. 
We consider the problem as to how accurately we can make the gate operation 
$\cE_{U,\rho_{ \bA }}$ to realize the quantum NOT  gate $\cE_{X_{ \bS}}$.
The worst error probability of this realization is defined by the
completely bounded distance \cite{Pau86,BDR05}
(the CB distance, or the half-CB-norm-distance) 
between $\cE_{U,\rho_{\bA}}$ and $\cE_{X_{\bS}}$, 
given by 
\begin{eqnarray}
\lefteqn{D_{{\mathrm{CB}}} (\cE_{U, \rho_{\bA}}, \cE_{X_{\bS}} )}\quad
\nonumber \\ 
&=&\sup_{n,\rho} D \left(  
\cE_{U,\rho_{\bA}}\otimes id_{n} \left( \rho \right), 
\cE_{X_{\bS}}  \otimes id_{n} \left( \rho \right) 
\right),  \label{D_CB()}
\end{eqnarray}
where $D(\cdot , \cdot)$ denotes the trace distance (or the half-trace-norm-distance) 
\cite{NC00}  defined by
$$
D(\rho_1,\rho_2)=\frac{1}{2}\Tr[|\rho_1-\rho_2|]
$$
for any states $\rho_1$ and $\rho_2$ of $\bS$,
$id_{n}$ is the identity operation on an $n$-level system ${\mathbf{E}}$, and 
$\rho$ runs over the density operators on $\bS + {\mathbf{E}}$. 
Since the trace distance of the output states can be interpreted 
as the achievable upper bound on the classical trace distances, or the
total-variation distances, between the probability distributions 
arising from any measurements on those states \cite{NC00}, 
the CB distance can be interpreted as the ultimate achievable upper bound 
on those classical trace distances with further allowing measurements 
over the environment with entangled input states;
see, for example, \cite{05AAE} for a discussion on
the enhancement of channel discriminations with
an entanglement assistance.
Thus, we interpret 
$D_{\mathrm{CB}}( \cE_{U, \rho_{\bA}}, \cE_{X_{\bS}} ) $
  as the worst error probability 
of $\cE_{U, \rho_{\bA}}$ in realizing
$\cE_{X_{\bS}}$. 
The  phrase ``error probability'' in the following discussion 
means the CB distance (\ref{D_CB()}).
Clearly, 
$$
D_{{\mathrm{CB}}} (\cE_{U, \rho_{\bA}}, \cE_{X_{\bS}} ) 
\geq \max_{\rho_{\bS}} D (   \cE_{U,\rho_{\bA}}  ( \rho_{\bS}) , 
\cE_{X_{\bS}}  ( \rho_{\bS} )),
$$
 and minimizing 
$\max_{\rho_{\bS}} D (\cE_{U,\rho_{\bA}}  ( \rho_{\bS} ) ,\cE_{X_{\bS}}  ( \rho_{\bS}))$
over
all the conservative implementations $(U,\rho_{\bA})$,
we find
\begin{eqnarray}
\lefteqn{D_{\mathrm{CB}} (  \cE_{U, \rho_{\bA}} ,{ \cE}_{X_{\bS}} ) }\quad
\nonumber \\
&\geq  & \min_{(U,\rho_{\bA})}\max_{\rho_{{\bS}}} 
D \left( \cE_{U, \rho_{\bA}}(\rho_{{\bS}}),\cE_{X_{\bS}}(\rho_{{\bS}})  \right).
\end{eqnarray}
The right-hand side of this inequality can be interpreted as a precision limit of the 
quantum NOT gate under the conservation law (\ref{c.l.}). 
If the limit could take 
zero,  it might be considered that there exists a perfect realization in $\cE_{U, \rho_{\bA}}$. 
However, we show that such a realization does not exist 
because of the conservation law (\ref{c.l.}). 

\subsection{Sufficiency of pure input states \label{max_input=pure-state}}

Now, we shall simplify the maximization over the input state $\rho_{\bS}$ 
by showing that it suffices to consider only pure state $\rho_{\bS}$.
To show this, we use the fact that the 
output trace distance is jointly convex in its inputs:  
\begin{eqnarray}
\label{convexity_D}
&&D \Big( \cE_{U,\rho_{\bA}} \Big( \sum_{i} p_{i}\rho_{i}\Big) ,
\cE_{X_{\bS}}\Big( \sum_{i} p_{i} \rho_{i} \Big) \Big) \nonumber \\  &&
\leq   
 \sum_{i} p_{i} D \left( \cE_ {U,\rho_{\bA}}\big( \rho_{i} \big) ,
\cE_{X_{\bS}}\big( \rho_{i} \big)  \right),
\end{eqnarray}
where $\sum_{i}p_{i}=1$ and $p_{i}\geq 0$.
This follows easily from the joint convexity of the trace distance \cite{NC00}
and the linearity of operations 
$\cE_{X_{\bS}}$ and $\cE_{U,\rho_{\bA}}$.

From the above inequality, a pure input state certainly gives the maximum of the trace distance.
To see this briefly, let 
$\rho_{\bS} = \sum_{i}q_{i} \left| \psi_{i} \right\rangle \left\langle \psi_{i} \right|$,
where 
$\sum_{i}q_{i}=1$ and $q_{i}\geq 0$.
Then, there exists a pure state 
$| \psi_{j} \rangle $ such that
\begin{eqnarray}
&& D \Big( \cE_{U,\rho_{\bA}} \Big( \sum_{i} q_{i} | \psi_{i} \rangle \langle \psi_{i} | \Big) ,
\cE_{X_{\bS}}\Big( \sum_{i} q_{i} \left| \psi_{i} \right\rangle \left\langle \psi_{i} \right| \Big)   \Big) 
\nonumber \\
& & \leq \sum_{i} q_{i} D ( \cE_{U,\rho_{\bA}}( | \psi_{i} \rangle \langle \psi_{i} | ) ,
\cE_{X_{\bS}}(   | \psi_{i} \rangle \langle \psi_{i} | )   ) 
\nonumber \\
&  & \leq D ( \cE_ {U,\rho_{\bA}}( | \psi_{j} \rangle \langle \psi_{j} | ) ,
\cE_{X_{\bS}} (   | \psi_{j} \rangle \langle \psi_{j} | )   ). 
\end{eqnarray}
Thus in considering $\max_{\rho_{{\bS}}} 
D ( \cE_{U, \rho_{\bA}}(\rho_{{\bS}}), \cE_{X_{\bS}}(\rho_{{\bS}})  ),$    
we shall assume in later discussions without loss of 
generality that the input state $\rho_{{\bS}}$ is a pure state. 

\subsection{Pure conservative implementations \label{se:pure_implementation}}

An implementation $(U,\rho_{\bA})$ is said to be pure if $\rho_{\bA}$ is a pure state.
In this case, we shall  write $(U,\rho_{\bA})=(U,\ket{A})$ if $\rho_{\bA}=\ketbra{A}$.
In the following sections, we shall mainly consider the case where the ancilla
state is a pure state.   
Here, we shall show a purification method that makes any general
conservative implementation a pure conservative implementation,
so that every conservative implementation with $N$ qubit ancilla has 
a pure conservative implementation with 
$N+\lceil\log_{2}\rank(\rho_{\bA})\rceil$ qubit
ancilla, where $\rank(\rho_{\bA})$ denotes the rank of $\rho_{\bA}$.

Let $(U,\rho_{\bA})$ be a conservative implementation 
with $N$ qubit ancilla $\bA$.
Then, we have the spectral decomposition
\beqa
\rho_{\bA}=\sum_{j=1}^{R}p_j\ketbra{\phi_j},
\eeqa
where $R=\rank(\rho_{\bA})$, $\bracket{\phi_j|\phi_k}=\delta_{jk}$,  
$p_j>0$ for all $j,k=1,\ldots,R$, and $\sum_{j}p_j=1$.
Let $\bA'$ be the $N'$ qubit ancilla system extending $\bA$
satisfying $N'=N+\lceil\log_2 R\rceil$.
Let $\ket{A'}\in\cH_{\bA'}$ be such that
\beqa\label{eq:Schmidt}
\ket{A'}=\sum_{j=1}^{R}\sqrt{p_j}\ket{\phi_j}\otimes\ket{\xi_j},
\eeqa
where $\ket{\xi_j}\in\cH_{\bA'-\bA}$, 
$\bracket{\xi_j|\xi_k}=\delta_{jk}$ for all $j,k=1,\ldots,R$.
We define a unitary operator $U'$ on $\cH_{\bS} \otimes \cH_{\bA}\otimes\cH_{\bA'-\bA}$ by
$U'=U\otimes I$, where $I$ is the identity operator on $\cH_{\bA'-\bA}$.

Now, we consider the implementation $(U',\ket{A'})$.
It is easy to see that $U'$ satisfies the conservation law $[U',Z]=0$,
where $Z$ is the sum of Pauli $Z$ operators in $\bS + \bA'$. 
We shall show the relation 
\beqa\label{eq:purification}
\cE_{U,\rho_{\bA}}=\cE_{U',\ket{A'}}.
\eeqa

Let $\rho_{\bS}$ be any input state. 
Then, by \Eq{eq:Schmidt} we have
\beqa\label{eq:purification2}
\Tr_{\bA'-\bA}[\rho_{\bS}\otimes\ketbra{A'}]=\rho_{\bS}\otimes\rho_{\bA}.
\eeqa
We also have
\beqas
\lefteqn{\cE_{U',\ket{A'}}(\rho_\bS)}\\
&=&
\Tr_{\bA'}[U'(\rho_{\bS}\otimes\ketbra{A'}){U'}^{\dagger}]\\
&=&
\Tr_{\bA}\Tr_{\bA'-\bA}[(U\otimes I)
(\rho_{\bS}\otimes\ketbra{A'}) (U^{\dagger}\otimes I)]\\ 
&=&
\Tr_{\bA}[U\Tr_{\bA'-\bA}[\rho_{\bS}\otimes\ketbra{A'}]U^{\dagger}].
\eeqas
From \Eq{eq:purification2}, we have
$$
\cE_{U',\ket{A'}}(\rho_\bS)=\Tr_{\bA}[U(\rho_{\bS}\otimes\rho_{\bA})U^{\dagger}].
$$
Since $\rho_{\bS}$ is arbitrary, \Eq{eq:purification} follows from \Eq{eq:operation}.

The implementation $(U',\ket{A'})$ is a pure conservative implementation 
and  has
$N'= N+\lceil\log_{2}\rank(\rho_{\bA})\rceil$ qubit ancilla.

\subsection{Gate fidelity and gate trace distance}

For any two trace-preserving CP maps $\cE_1$ and $\cE_2$ their distance measures
are defined as follows.  The gate fidelity \cite{NC00}
$F(\cE_1,\cE_2)$ between $\cE_1$ and $\cE_2$  is defined by 
\beqa\label{eq:gate_fidelity}
F(\cE_1,\cE_2)
=\inf_{\rho_\bS}F(\cE_1(\rho_\bS),\cE_2(\rho_\bS)),
\eeqa
where $\rho_\bS$ varies over all the states of $\bS$ and $F(\cdot,\cdot)$ in the right-hand-side 
denotes the
fidelity defined by
\beqa
F(\rho_1,\rho_2)=\Tr[(\rho_1^{1/2}\rho_2\rho_1^{1/2})^{1/2}]
\eeqa
for all states $\rho_1$ and $\rho_2$ of $\bS$.  
By the joint concavity of the fidelity \cite[p.~415]{NC00} the infimum in
\Eq{eq:gate_fidelity} 
can be replaced by the one over only all the pure states $\rho_{\bS}$ of $\bS$.

We define
the gate trace-distance $D(\cE_1,\cE_2)$ between $\cE_1$ and $\cE_2$  by
\beqa\label{eq:gate_trace_distance}
D(\cE_1,\cE_2)=\sup_{\rho_\bS}D(\cE_1(\rho_\bS),\cE_2(\rho_\bS)),
\eeqa
where $\rho_\bS$ varies over all the states of $\bS$.
By the result obtained in subsection \ref{max_input=pure-state},
the supremum in \Eq{eq:gate_trace_distance}
can be replaced by the one over only all the pure states $\rho_{\bS}$ of $\bS$.

For any state $\rho_1$ and any pure state $\rho_2$, 
the fidelity and the trace distance are related by
\beqas
D(\rho_1,\rho_2)\ge 1-F(\rho_1,\rho_2)^{2}
\eeqas
(see Eq.~(9,111) of Ref.~\cite{NC00}).
Since $\cE_{X_{\bS}}(\rho_{\bS})$ is a pure state provided that $\rho_{\bS}$ is pure,
we have
\beqa\label{eq:trace_d<infidelity}
D(\cE_{U,\rho_{\bA}}(\rho_{\bS}),\cE_{X_{\bS}}(\rho_{\bS}))
\ge 1-F(\cE_{U,\rho_{\bA}}(\rho_{\bS}),\cE_{X_{\bS}}(\rho_{\bS}))^{2}\nonumber\\
\eeqa
for any pure state $\rho_{\bS}$ of $\bS$.
Taking supremum over all the pure states $\rho_{\bS}$ 
of the both sides of \Eq{eq:trace_d<infidelity}, for any  implementation $(U,\rho_{\bA})$ 
we obtain
\begin{eqnarray}
D_{{\mathrm{CB}}}(\cE_{U,\rho_{\bA}},\cE_{X_{\bS}})
& \ge& D(\cE_{U,\rho_{\bA}},\cE_{X_{\bS}})\nonumber\\
&\ge& 1-F(\cE_{U,\rho_{\bA}},\cE_{X_{\bS}})^{2}.
\end{eqnarray}

In Ref.~\cite{03UPQ}, the realization of the Hadamard gate
$H_{\bS}=(1/\sqrt{2})(\ket{0}\bra{0}+\ket{1}\bra{0}+\ket{0}\bra{1}-\ket{1}\bra{1})$ 
has been considered and  it has
been proved that for any pure conservative implementation
$(U,\ket{A})$ with $N$ qubit ancilla $\bA$, we have
\begin{equation}\label{eq:Hadamard}
1-F(\cE_{U,\ket{A}},\cE_{H_{\bS}})^{2}\ge
\frac{1}{4N^2+4},
\end{equation}
where $\cE_{H_{\bS}}(\rho_{\bS})=H_{\bS}\rho_{\bS}H_{\bS}^{\dagger}$
 \footnote{Note that the presentation of Ref.~\cite{03UPQ} discusses the
conservation law for the $x$-component of the spin instead of the $z$-component 
considered in the present paper. However, in that argument the $x$-component and the
$z$-component are completely interchangeable, since we have 
both relations $H^{\dagger}XH=Z$ and 
$H^{\dagger}ZH=X$
from $H=H^{\dagger}$.}.
Since any conservative implementation  $(U,\rho_{\bA})$ with $N$ qubit ancilla $\bA$
can be purified to be a pure conservative implementation  $(U',\ket{A'})$ 
with $N+\lceil\log_2\rank(\rho_{\bA})\rceil$ qubit ancilla $\bA'$, we have
\beqa\label{eq:Hadamard2}
1-F(\cE_{U,\rho_{\bA}},\cE_{H_{\bS}})^{2}\ge
\frac{1}{4(N+\log_2\rank(\rho_{\bA}))^2+4}.
\eeqa
Since $N+\lceil\log_2\rank(\rho_{\bA})\rceil\le2N$, we conclude that every 
conservative implementation  $(U,\rho_{\bA})$ with $N$
qubit ancilla $\bA$ satisfies
\beqa\label{eq:Hadamard3}
1-F(\cE_{U,\rho_{\bA}},\cE_{H_{\bS}})^{2}\ge
\frac{1}{16N^2+4}.
\eeqa
In other words, we have
\beqa
\min_{(U,\ket{A})}\max_{\rho_{\bS}}[1-F(\cE_{U,\ket{A}},\cE_{H_{\bS}})^{2}]
\ge \frac{1}{4N^2+4},
\eeqa
where $(U,\ket{A})$ varies over all the pure conservative implementations with $N$ qubit
ancilla $\bA$, and we have
\beqa
\min_{(U,\rho_{\bA})}\max_{\rho_{\bS}}[1-F(\cE_{U,\rho_{\bA}},\cE_{H_{\bS}})^{2}]
\ge \frac{1}{16N^2+4},
\eeqa
where $(U,\rho_{\bA})$ varies over all the conservative implementations with $N$ qubit
ancilla $\bA$.

\subsection{Main results}

Unfortunately, the method for deriving \Eq{eq:Hadamard} cannot be applied to 
the quantum NOT gate. In this paper we develop a new method for analyzing 
the gate trace-distance  $D(\cE_{U,\rho_{\bA}},\cE_{X_{\bS}})$ instead of considering the gate
fidelity $F(\cE_{U,\rho_{\bA}},\cE_{X_{\bS}})$ and we shall prove 
the following relations.  In section \ref{accuracy-limit-size}, we shall show that any 
pure conservative implementation $(U,\ket{A})$ with $N$ qubit ancilla satisfies
\begin{eqnarray}\label{eq:pci}
D (\cE_{U,\ket{A}},\cE_{X_{\bS}} ) 
\geq  \frac{1}{2}
\Big( 1 -   \cos\frac{2 \pi}{  N+4  } \Big).
\end{eqnarray}
It follows from the above, any conservative implementation $(U,\rho_{\bA})$ 
with $N$ qubit ancilla satisfies
\begin{eqnarray}\label{eq:ci}
D (\cE_{U,\rho_{\bA}},\cE_{X_{\bS}} ) 
\geq  \frac{1}{2}
\Big( 1 -   \cos  \frac{2 \pi}{  N+\log_2 \rank(\rho_{\bA})+4 } \Big).\quad
\end{eqnarray}
An implementation $(U,\rho_{\bA})$ is called a classically complete implementation
of the quantum NOT gate, or classically complete implementation for short,
if it satisfies
\begin{eqnarray}
\cE_{U,\rho_{\bA}}(\ketbra{0})&=&\ketbra{1},\\
\cE_{U,\rho_{\bA}}(\ketbra{1})&=&\ketbra{0}.
\end{eqnarray}
In section \ref{a-noteworthy}, we shall consider classically complete pure
conservative implementations.
We shall find the attainable lower bound for this case, so that we obtain
\beqa\label{eq:ccpcie}
\lefteqn{
\min_{(U,\ket{A})}\max_{\rho_{\bS}}
D (\cE_{U,\left| A \right\rangle}( \rho_{\bS}),\cE_{X_{\bS}} ( \rho_{\bS} )  ) }\quad
\nonumber\\
&=& \frac{1}{2}\Big(   1-  \cos \frac{2\pi}{N +2}\Big),
\eeqa
where $(U,\ket{A})$ varies over all the classically complete pure
conservative implementations with $N$ qubit ancilla $\bA$ provided $N$ is even, 
and we obtain
 \beqa\label{eq:ccpcio}
\lefteqn{
\min_{(U,\ket{A})}\max_{\rho_{\bS}}
D (\cE_{U,\left| A \right\rangle} ( \rho_{\bS} ) , \cE_{X_{\bS}}( \rho_{{\bS}}) ) }\quad
\nonumber\\
&=&
\frac{1}{2}\Big( 1-  \cos \frac{2\pi}{N +1} \Big),
\eeqa
provided $N$ is odd.  
From the above, any classically complete
conservative implementation $(U,\rho_{\bA})$ 
with $N$ qubit ancilla satisfies
\begin{eqnarray}
D (\cE_{U,\rho_{\bA}},\cE_{X_{\bS}} ) 
\geq  \frac{1}{2}
\Big( 1 -   \cos  \frac{2 \pi}{  N+\log_2 \rank(\rho_{\bA})+2} \Big).\quad
\end{eqnarray}

Since $N+\log_2 \rank(\rho_{\bA})\le 2N$, from the above we have
\beqa
\lefteqn{\frac{1}{2}\Big(   1-  \cos \frac{2\pi}{N +1}\Big)}\quad\nonumber\\
&\ge&
\min_{(U,\rho_{\bA})}\max_{\rho_{\bS}}
D(\cE_{U,\rho_{\bA}}(\rho_\bS),\cE_{X_{\bS}}(\rho_\bS))\nonumber\\
&\ge&
 \frac{1}{2}\Big(   1-  \cos \frac{\pi}{N+1}\Big),
\eeqa
where $(U,\rho_{\bA})$ varies over all the classically complete implementations
with $N$ qubit ancilla.
From Eqs.~(\ref{eq:pci}) and (\ref{eq:ccpcio}), we have
\begin{eqnarray}
\lefteqn{\frac{1}{2}\Big(   1-  \cos \frac{2\pi}{N +1}\Big)}\quad\nonumber\\
&\ge&
\min_{(U,\ket{A})}
\max_{\rho_{{\bS}}}
D (\cE_{U,\ket{A}}( \rho_{{\bS}}),\cE_{X_{\bS}} 
( \rho_{\bS} ) ) \nonumber\\
&\geq&  \frac{1}{2}
\Big(1 -   \cos \frac{2 \pi}{  N+4  } \Big),
\end{eqnarray}
where $(U,\ket{A})$ varies over all the pure conservative implementations.
Finally,  from Eqs.~(\ref{eq:ci}) and  (\ref{eq:ccpcio}), we have
\begin{eqnarray}
\lefteqn{\frac{1}{2}\Big(   1-  \cos \frac{2\pi}{N +1}\Big)}\quad\nonumber\\
&\ge&
\min_{(U,\rho_{\bA})}\max_{\rho_{\bS}}
D (\cE_{U,\rho_{\bA}}( \rho_{{\bS}}),\cE_{X_{\bS}} 
( \rho_{\bS} ) ) \nonumber\\
&\geq&  \frac{1}{2}
\Big( 1 -   \cos \frac{\pi}{  N+2 } \Big),
\end{eqnarray}
where $(U,\rho_{\bA})$ varies over all the conservative implementations
with $N$ qubit ancilla $\bA$.

\section{Lower bound of gate trace distance \label{Der-of-the-}}

In this section, we investigate the maximum trace distance over all possible input states of $\bS$
for given $U$ and $\rho_{\bA}$ in a general way without considering the
conservation law. 

\subsection{System input state and  trace distance}

We start with a description of the output states controlled by any
unitary operator $U$ on $\cH_{\bS}\otimes\cH_{\bA}$. 
Any pure input state $\ket{\psi}$ of $\bS$ can be
described as
\begin{eqnarray}
\left| \psi \right\rangle = \alpha \ket{0} + \beta \ket{1},
\end{eqnarray}
where $| \alpha |^{2} + | \beta |^{2}=1$.
We assume that the input state of $\bA$ is a pure state $\ket{A}$, so 
that the input state of the composite system $\bS+\bA$ is 
the product state  $\ket{\psi} \otimes \ket{A} $. When $\ket{0}$ 
or  $\ket{1}$ is an input state of $\bS$ 
the corresponding output state of $\bS+\bA$ can be generally expressed as 
\begin{eqnarray}
U \left( \ket{0} \otimes \left| A \right\rangle \right)
&=& \ket{0}  \otimes \ket{A^{0}_{0}}  
+ \ket{1}  \otimes \ket{A^{0}_{1}},  \nonumber\\
U \left( \ket{1} \otimes \ket{A} \right)
&=& \ket{0}  \otimes \ket{A^{1}_{0}}  
+ \ket{1}  \otimes \ket{A^{1}_{1}},  \label{U0U1}  
\end{eqnarray}
where 
 $ | A^{i}_{j} \rangle \in \cH_{\bA}$ for $i,j=0,1$. 
Normalizing these states gives
\begin{eqnarray}
\| \ket{A^{0}_{0}}  \|^{2}
+\| \ket{A^{0}_{1}}  \|^{2} &=&1, \nonumber \\
\|  \ket{A^{1}_{0}}  \|^{2}
+\|  \ket{A^{1}_{1}}  \|^{2} &=&1.
\label{Nor.con.Aij}
\end{eqnarray}
The output state of $\bS+\bA$ corresponding to 
$\ket{\psi} $  can then be expressed as 
\begin{eqnarray}
U ( \ket{\psi} \otimes \ket{A}   )
&=& \alpha \left( \ket{0}  \otimes \ket{A^{0}_{0}}  
+ \ket{1}  \otimes \ket{A^{0}_{1}} \right)  \nonumber \\
&&+ \beta \left( \ket{0}  \otimes \ket{A^{1}_{0}}  
+ \ket{1}  \otimes \ket{A^{1}_{1}}  
\right) .
\label{U(psi_A)general}
\end{eqnarray}
Normalizing Eq.~(\ref{U(psi_A)general}) gives
\begin{eqnarray}
{\mathrm{Re}}\left[ \alpha^{*}\beta  \left( 
\langle A^{0}_{0} \ket{A^{1}_{0}} 
+\langle A^{0}_{1} \ket{A^{1}_{1}} 
\right)\right] =0. \label{Nor.con.A}
\end{eqnarray}
The output state  $\cE_{U, \ket{A}}( \left| \psi \right\rangle )
:=\cE_{U, \ket{A}}( | \psi\rangle\langle\psi| )$
of $\bS$ is 
given by the partial trace  of Eq. (\ref{U(psi_A)general}) with respect to $\bA$
as follows.
\begin{widetext}
\begin{eqnarray}
\cE_{U, \ket{A}}( \left| \psi \right\rangle )
&=&{\mathrm{Tr}}_{\bA} 
[ U ( \ket{\psi}\langle  \psi | \otimes \ket{A}\langle A | )U^{\dagger} ] \nonumber \\
&=&( |\alpha|^{2} \| | A^{0}_{0} \rangle  \|^{2}
+\alpha \beta^{*}  \langle A^{1}_{0} \ket{A^{0}_{0}} 
+\alpha^{*}\beta  \langle A^{0}_{0} \ket{A^{1}_{0}} 
+|\beta |^{2}   \| | A^{1}_{0} \rangle  \|^{2} )   
\ket{0}\left\langle  0 \right|  \nonumber\\
& &+ \left( |\alpha |^{2} \langle A^{0}_{1} \ket{A^{0}_{0}}
+\alpha \beta^{*}\langle A^{1}_{1} \ket{A^{0}_{0}} 
+\alpha^{*}\beta  \langle A^{0}_{1} \ket{A^{1}_{0}}
+|\beta |^{2}  \langle A^{1}_{1} \ket{A^{1}_{0}}  
\right) \ket{0}\left\langle  1 \right|\nonumber\\
& & + \left( |\alpha |^{2} \langle A^{0}_{0}\ket{A^{0}_{1}}  
+\alpha \beta^{*}\langle A^{1}_{0}\ket{A^{0}_{1}} 
+\alpha^{*}\beta  \langle A^{0}_{0} \ket{A^{1}_{1}}
+|\beta |^{2}  \langle A^{1}_{0} \ket{A^{1}_{1}}  
\right) \ket{1}\left\langle  0 \right|\nonumber\\
& &+ \left( |\alpha |^{2} \|\ket{A^{0}_{1}}  \|^{2}
+\alpha \beta^{*} \langle A^{1}_{1}\ket{A^{0}_{1}} 
+\alpha^{*}\beta  \langle A^{0}_{1} \ket{A^{1}_{1}}
+|\beta |^{2} \| | A^{1}_{1} \rangle  \|^{2}
\right) \ket{1}\left\langle  1 \right| .
\label{E_U,|A>=Tr_A[U(product)U]}
\end{eqnarray}
On the other hand, if the quantum NOT gate were to be perfectly realized,
the output state $\cE_{X_{{\bS}}}( \left| \psi \right\rangle )
:=\cE_{X_{{\bS}}}( |\psi\rangle\langle\psi|)$  would be given by
\begin{eqnarray}
\cE_{X_{{\bS}}}( \left| \psi \right\rangle ) 
= X_{{\bS}} \left| \psi \right\rangle \left\langle \psi \right|  X_{{\bS}}^{\dagger} 
= |\beta |^{2}  \ket{0}\left\langle  0 \right| 
+ \alpha^{*} \beta \ket{0}\left\langle  1 \right|
+ \alpha \beta^{*} \ket{1}\left\langle  0
\right|  + |\alpha |^{2} \ket{1}\left\langle  1 \right|. 
\label{rhoXij}
\end{eqnarray}
We now consider the trace distance between $\cE_{U, \ket{A}}( \ket{\psi} )$ and 
$\cE_{X}( \ket{\psi} ).$
Note that the trace distance between two-dimensional states, $\sigma^{\xi}$ and $\sigma^{\eta}$,  
can be described as 
\begin{eqnarray}
D( \sigma^{\xi},  \sigma^{\eta}) 
= 
\sqrt{|\sigma^{\xi}_{01}- \sigma^{\eta}_{01}|^{2}
-(\sigma^{\xi}_{00}- \sigma^{\eta}_{00})(\sigma^{\xi}_{11}- \sigma^{\eta}_{11})}, 
\label{T-D-component}
\end{eqnarray}
where 
$\sigma_{ij}^{k}= \langle i| \sigma^{k} | j \rangle $ for $k=\xi,\eta$.
Using this relation,  
the trace distance $D(\cE_{X_{{\bS}}}( \left| \psi \right\rangle),
 \cE_{U,\ket{A}}( \left| \psi \right\rangle ))$
is 
\begin{eqnarray}
\lefteqn{D(\cE_{U,\ket{A}}( \left| \psi \right\rangle) ,
 \cE_{X_{{\bS}}}( \left| \psi \right\rangle ))}\quad
\nonumber \\
&= &
\Big\{
          \big|\alpha^{*}\beta -  
                 \big( |\alpha |^{2}\langle A^{0}_{1} \ket{A^{0}_{0}}
                        +\alpha \beta^{*}\langle A^{1}_{1} \ket{A^{0}_{0}} 
                        +\alpha^{*}\beta  \langle A^{0}_{1} \ket{A^{1}_{0}}
                        +|\beta |^{2}  \langle A^{1}_{1} \ket{A^{1}_{0}}  
                \big) 
         \big|^{2} 
\nonumber \\
& &
- \big[ |\beta |^{2}
        - \big( |\alpha|^{2} \| | A^{0}_{0} \rangle \|^{2}
                 +\alpha \beta^{*}  \langle A^{1}_{0} \ket{A^{0}_{0}} 
                 +\alpha^{*}\beta  \langle A^{0}_{0} \ket{A^{1}_{0}} 
                 +|\beta |^{2}   \| | A^{1}_{0} \rangle \|^{2}
         \big) 
  \big]
\nonumber \\
&&
\times \big[ |\alpha |^{2} 
                 -\big( |\alpha |^{2} \|\ket{A^{0}_{1}}  \|^{2}
                          +\alpha \beta^{*} \langle A^{1}_{1}\ket{A^{0}_{1}} 
                          +\alpha^{*}\beta  \langle A^{0}_{1} \ket{A^{1}_{1}}
                          +|\beta |^{2} \| | A^{1}_{1} \rangle \|^{2}
                  \big) 
          \big] 
\Big\}^{\frac{1}{2}}. 
\label{D_incomplete}
\end{eqnarray}
Let  $ \epsilon_{0}=\|  | A^{0}_{0} \rangle  \|^{2}  $  
and $ \epsilon_{1} =\| | A^{1}_{1} \rangle  \|^{2}  $. 
Then $\|\ket{A^{0}_{1}}  \|^{2}  =1- \epsilon_{0}  $ and 
$\|  |A^{1}_{0} \rangle  \|^{2}  =1- \epsilon_{1} $ by Eq.~(\ref{Nor.con.Aij}). 
Thus  Eqs.~(\ref{Nor.con.A}) and (\ref{D_incomplete}) give 
\begin{eqnarray}
D(\cE_{U,\ket{A}}( \ket{\psi}) , \cE_{X_{{\bS}}}( \ket{\psi} ))
&= &\Big\{ \big| \alpha^{*}\beta ( 1- \langle A^{0}_{1} \ket{A^{1}_{0}} )
- \alpha \beta^{*}  \langle A^{1}_{1} \ket{A^{0}_{0}}  
- |\alpha |^{2} \langle A^{0}_{1} \ket{A^{0}_{0}}
 -|\beta |^{2}  \langle A^{1}_{1} \ket{A^{1}_{0}}
\big|^{2}  \nonumber\\ 
& &
+  \big[   -|\alpha|^{2} \epsilon_{0} + |\beta|^{2} \epsilon_{1}  
- 2 \mathrm{Re}\left( \alpha^{*}\beta   \langle A^{0}_{0} \ket{A^{1}_{0}}   
  \right) \big]^{2}  \Big\}^{\frac{1}{2}}.
  \label{D()_expre_as_epsilon}
\end{eqnarray}
Clearly
$\left[ \left(  -|\alpha|^{2} \epsilon_{0} + |\beta|^{2} \epsilon_{1} \right) - 2 \mathrm{Re}\left( 
\alpha^{*}\beta   \langle A^{0}_{0} \ket{A^{1}_{0}}   
  \right) \right]^{2} \geq 0 $,  and hence we obtain
\begin{eqnarray}
\lefteqn{D(\cE_{U,\ket{A}}( \ket{\psi}) , \cE_{X_{{\bS}}}( \ket{\psi} )) }\quad\nonumber \\
&\geq &
\big| \alpha^{*}\beta \left( 1- \langle A^{0}_{1} \ket{A^{1}_{0}} \right)
- \alpha \beta^{*}  \langle A^{1}_{1} \ket{A^{0}_{0}} 
 - |\alpha |^{2} \langle A^{0}_{1} \ket{A^{0}_{0}}\nonumber \\ 
 & &-|\beta |^{2}  \langle A^{1}_{1} \ket{A^{1}_{0}}
\big|.
\label{general-D(X,U)}
\end{eqnarray}
\end{widetext}

\subsection{Lower bound for maximum trace distance}

In the following, we shall prove that for any $U$ and $\ket{A}$,
we have
\begin{eqnarray}
\max_{\rho_{{\bS}}}
D \left( \cE_{U,\left| A \right\rangle}( \rho_{{\bS}} ) , \cE_{X_{\bS}}( \rho_{{\bS}})\right)
 & \geq & \frac{1}{2}\left|   1- \langle A^{0}_{1} \ket{A^{1}_{0}}  \right|,\quad
\label{general_inequality}
\end{eqnarray}
by considering the maximization of Eq.~(\ref{general-D(X,U)}) 
over the input state $\ket{\psi}$ of $\bS$.  
This means that the output trace distance 
must satisfy Eq.~(\ref{general_inequality}) for any interaction 
and any input state of $\bA$.

The proof is as follows.
We consider the input state $\ket{\psi'}=\alpha\ket{0}+\beta\ket{1}$
such that $|\alpha|=|\beta|=\frac{1}{\sqrt{2}}$. 
Let  $\theta$ be such that $\alpha^{*}\beta=\frac{1}{2}e^{i\theta}$ and $0 \leq \theta <
2\pi$. Then Eq.~(\ref{general-D(X,U)}) gives  
\begin{eqnarray}
\lefteqn{D(
\cE_{U,\ket{A}}( | \psi' \rangle ),\cE_{X_{{\bS}}}( | \psi' \rangle) ) }\quad
\nonumber \\
&\geq&  
\frac{1}{2}\big| e^{i\theta} \left( 1- \langle A^{0}_{1} \ket{A^{1}_{0}} \right)
- e^{-i\theta}  \langle A^{1}_{1} \ket{A^{0}_{0}}  
-  \langle A^{0}_{1} \ket{A^{0}_{0}}\nonumber \\
& & -  \langle A^{1}_{1} \ket{A^{1}_{0}}
\big|. 
\end{eqnarray}
Here three complex numbers, $1- \langle A^{0}_{1} \ket{A^{1}_{0}}$, 
$-\langle A^{1}_{1} \ket{A^{0}_{0}}$, and 
$- \langle A^{0}_{1} \ket{A^{0}_{0}} -  \langle A^{1}_{1} \ket{A^{1}_{0}}$, 
which are determined by $U$ and $\ket{A}$,
can be expressed as
\begin{eqnarray}
1- \langle A^{0}_{1} \ket{A^{1}_{0}} &=& r_{1}e^{i\phi_{1}}, \nonumber  \\
-\langle A^{1}_{1} \ket{A^{0}_{0}}  &=& r_{2}e^{i \phi_{2}}, \nonumber \\
 - \langle A^{0}_{1} \ket{A^{0}_{0}}
 -  \langle A^{1}_{1} \ket{A^{1}_{0}} &=&  r_{3}e^{i\phi_{3}},
\end{eqnarray}
where $r_{i} \geq 0$ and $0 \leq \phi_{i} < 2\pi$ for $i=1,2,3$. Then
$r_1=|   1- \langle A^{0}_{1} \ket{A^{1}_{0}} |$ and
\begin{eqnarray}
\lefteqn{D(\cE_{U,\ket{A}}( | \psi' \rangle ),
\cE_{X_{{\bS}}}( | \psi' \rangle) ) }\quad
\nonumber \\
&\geq&  \frac{1}{2}\big|  r_{1}
+  r_{2}e^{i(-2\theta - \phi_{1}+\phi_{2})} 
+r_{3}e^{i (-\theta +\phi_{3}- \phi_{1})}
\big|.
\label{r-D(X,U)}
\end{eqnarray}
Note that 
Eq.~(\ref{r-D(X,U)}) is maintained for any $\theta$ which is independent of $U$ and $\ket{A}$. 
Hence, we consider the following two cases. 
In the first case, suppose that $U$ and $\ket{A}$ satisfy $r_{2} \geq r_{3}$.  
In this case,  for the input state $ | \psi'_{a} \rangle$ of $\bS$ with
$\theta = \frac{1}{2}(\phi_{2}-\phi_{1})$, we have
\begin{eqnarray*}
\lefteqn{D(\cE_{U,\ket{A}}( | \psi'_{a} \rangle ),
\cE_{X_{{\bS}}}( | \psi'_{a} \rangle) 
) }\quad
\nonumber \\
&=& \frac{1}{2}
\big|  r_{1} +  r_{2} +r_{3}e^{i \{-\frac{1}{2}(\phi_{2}-\phi_{1}) +\phi_{3}- \phi_{1}\}} \big| 
\nonumber \\
&\geq  &\frac{1}{2}\left|  r_{1}
+  r_{2} - r_{3}
\right| \nonumber \\
&\geq  &\frac{1}{2} r_{1}. 
\end{eqnarray*}
Thus, there exists a state $\ket{\psi}$ of 
$\bS$ that satisfies $D(\cE_{X_{{\bS}}}(\ket{\psi})
, \cE_{U,\ket{A}}(\ket{\psi})) 
\geq \frac{1}{2}r_1$ 
in the case where $r_{2} \geq r_{3}$. 
In the second case,  suppose that $U$ and $\ket{A}$ satisfy
$r_{2} < r_{3}$.  In this case,   for the input state  $| \psi'_{b} \rangle$ with $ \theta =
\phi_{3}-\phi_{1}$, we have
\begin{eqnarray*}
\lefteqn{D(\cE_{U,\ket{A}}( | \psi'_{b} \rangle ),
\cE_{X_{{\bS}}}( | \psi'_{b} \rangle) ) }\quad
\nonumber\\
&=& \frac{1}{2}\big|  r_{1} +  r_{2} e^{i\{- 2(\phi_{3}-\phi_{1}) - \phi_{1}+\phi_{2}\}} +r_{3} \big| 
\nonumber \\
&\geq &  \frac{1}{2}\big|  r_{1} +  r_{3} - r_{2} \big| \nonumber \\
&\geq & \frac{1}{2} r_{1}. 
\end{eqnarray*}
Thus, there exists a state $\ket{\psi}$ of 
$\bS$ that satisfies $D( \cE_{U,\ket{A}}(\ket{\psi}),\cE_{X_{{\bS}}}(\ket{\psi})
)\geq  \frac{1}{2}r_{1}$ in the case where $r_{2} < r_{3}$.
We therefore conclude that for any $U$ and $\ket{A}$,
there exists a state $\ket{\psi}$ of 
 $\bS$ such that the input state $\rho_{\bS}=\ket{\psi}\bra{\psi}$
satisfies 
\beqa\label{eq:bound}
D( \cE_{U,\ket{A}}(\rho_{\bS}),\cE_{X_{{\bS}}}(\rho_{\bS})
) 
\geq \frac{1}{2}|   1- \langle A^{0}_{1} \ket{A^{1}_{0}}|.
\eeqa  
This completes the proof.

In Eq.~(\ref{general_inequality}), if the inner product $\langle A^{0}_{1} \ket{A^{1}_{0}}$ 
could take one by a certain choice of  $U$ and  $\ket{A}$, 
the lower bound could take zero. 
This may mean a perfect realization of $\cE_{U,\ket{A}}$ exists.  
However, we will show in the following sections that the inner product cannot take 
one by assuming the conservation law (\ref{c.l.}).
This result will give us a precision limit of the quantum NOT gate.

\section{Precision limit given the ancilla state \label{lowerbound-conservation}}

In this section, we derive the lower bound which depends on the input state of the ancilla system
by minimizing the right-hand-side of \Eq{eq:bound}
over the evolution operator $U$  under the conservation law.

\subsection{Constraints on ancilla input states}
 
We start with the description of the input state of $\bA$. 
The sum of the Pauli Z operators on $\bA$ is the operator $Z_{\bA}$
on $\cH_{\bA}$ given by
$$
Z_{\bA}=\sum_{i=1}^{N} {Z}_{{\bA}_{i}}.
$$
We denote the eigenspace in $Z_{\bA}$ of an eigenvalue $\xi$ by $E_{\xi}^{Z_{\bA}}$.
The eigenvalues are $N-2n$, where $n = 0,1,2,\cdots,N$.  
The dimension of the eigenspace  of the eigenvalue $N-2n$ is $d_n=\frac{N!}{(N-n)!n!}$. 
Note that the Hilbert space of $\bA$ is the direct sum 
of the spaces $E_{N-2n }^{Z_{\bA}}$ for
$n=0,1,\cdots,N$:
\begin{eqnarray}
\cH_{\bA}= \oplus_{n = 0}^{N} E_{N-2n }^{Z_{\bA}}.
\label{H_A=o+E^A}
\end{eqnarray}
Therefore, for any input state $\ket{A}$ of $\bA$ there exist 
 $a_{n}\in {\mathbf{C}}$  and 
$ | \phi_{n}^{A}  \rangle \in E_{N-2n }^{Z_{\bA}}$ with
 $\|  | \phi_{n}^{A}  \rangle  \|=1$ satisfying
\begin{equation}
\ket{A} 
= \sum_{  n = 0  }^{N} a_{n} | \phi_{n}^{A}  \rangle .  \hspace{3mm}
\label{A=sum_a_nn}
\end{equation}
Normalizing Eq.~(\ref{A=sum_a_nn}) gives
\begin{eqnarray}\label{sum_an}
\sum_{n=0}^{N} |a_{n}|^{2}=1.
\label{nor-con-a_n}
\end{eqnarray}

Next we describe the output state of $\bS+\bA$ under the conservation law.
Let $E_{m}^{ { Z_{{\bS}}}}$ be 
the eigenspace of an eigenvalue $m=1,-1$ of  $Z_{{\bS}}$,
and $ E_{\lambda}^{Z}$ be the eigenspace of  an eigenvalue $ \lambda$ of $Z$,
where $Z= {Z}_{{\bS}}+{Z}_{\bA}$, 
which has
\begin{eqnarray}
\lambda=N+1-2n,
\end{eqnarray}
where $n=0,1,\cdots, N,N+1$.
The eigenspace $ E_{\lambda}^{Z}$ can be expressed 
by the tensor product of the space  $  E_{1}^{ {
Z_{{\bS}}}}\otimes E_{ N-2n}^{ { Z_{\bA}} } $ 
and the space
$  E_{-1}^{ { Z_{{\bS}}}}\otimes E_{ N-2n}^{
{ Z_{\bA}} } $ as follows:
\begin{eqnarray}
E_{N+1}^{Z} &=& E_{1}^{ { Z_{{\bS}}}} \otimes E_{ N }^{ \mathrm{ Z_{\bA}} }, \nonumber \\
E_{N+1-2}^{Z}&=& \big(  E_{1}^{ { Z_{{\bS}}}} \otimes E_{ N-2 }^{ \mathrm{ Z_{\bA}} } \big)
\oplus  \big( E_{-1}^{ { Z_{{\bS}}}}\otimes E_{ N }^{ \mathrm{ Z_{\bA}} } \big),
\nonumber \\
E_{N+1-4}^{Z}&=& \big( E_{1}^{ { Z_{{\bS}}}}\otimes E_{ N-4 }^{ \mathrm{ Z_{\bA}} } \big)
 \oplus \big( E_{-1}^{ { Z_{{\bS}}}} \otimes E_{ N-2 }^{ \mathrm{ Z_{\bA}} } \big),
\nonumber \\
&\vdots&  
\nonumber \\
E_{N+1-2n}^{Z}&=&  \big( E_{1}^{ { Z_{{\bS}}}} \otimes E_{N-2n }^{ \mathrm{ Z_{\bA}} } \big)
\oplus \big(  E_{-1}^{ { Z_{{\bS}}}} \otimes E_{ N-2(n-1) }^{ \mathrm{ Z_{\bA}} } \big),
\nonumber \\
&\vdots&  
\nonumber \\
E_{ -N + 1 }^{Z}&=& \big( E_{1}^{ { Z_{{\bS}}}} \otimes E_{ -N  }^{ \mathrm{ Z_{\bA}} } \big) 
\oplus  \big( E_{-1}^{ { Z_{{\bS}}}} \otimes E_{ -N+2 }^{ \mathrm{ Z_{\bA}} } \big),
\nonumber \\
E_{ -N - 1 }^{Z}&=&   E_{-1}^{ { Z_{{\bS}}}} \otimes E_{ - N }^{ \mathrm{Z_{\bA}} }.
\label{E=E+E}
\end{eqnarray}

Note that the conservation law (\ref{c.l.}) can be equivalently expressed by the
relation \footnote{
To see this, let $P_{\lambda}$ the projection on $E_{\lambda}^{Z}$.
Then, (\ref{[U,Z]=0arrowUEsubsetE}) is equivalent to 
$UP_{\lambda}=P_{\lambda}UP_{\lambda}$ for all $\lambda$, whereas
(\ref{c.l.}) is equivalent to $UP_{\lambda}=P_{\lambda}U$ for all $\lambda$.
Thus, (\ref{c.l.}) implies (\ref{[U,Z]=0arrowUEsubsetE}).
Conversely, from (\ref{[U,Z]=0arrowUEsubsetE}) we also have
$U(I-P_{\lambda})=(I-P_{\lambda})U(I-P_{\lambda})$ to obtain 
$P_{\lambda}U=P_{\lambda}UP_{\lambda}$ for all $\lambda$, 
and consequently (\ref{c.l.}) follows from (\ref{[U,Z]=0arrowUEsubsetE}).}
\begin{eqnarray}
UE_{\lambda}^{Z}  \subset E_{\lambda}^{Z} 
\label{[U,Z]=0arrowUEsubsetE}
\end{eqnarray}
for all $\lambda$.
Eqs.~(\ref{E=E+E}) and (\ref{[U,Z]=0arrowUEsubsetE}) then show that 
the output state $U (  \ket{0} \otimes |  \phi_{n}^{A} \rangle )$ 
is an element of the subspace 
$   ( E_{1}^{ { Z_{{\bS}}}} \otimes E_{N-2n }^{ \mathrm{ Z_{\bA}} } )
\oplus (  E_{-1}^{ { Z_{{\bS}}}} \otimes E_{ N-2(n-1) }^{ \mathrm{ Z_{\bA}} } )  $ 
for $n=1,2,\cdots,N$, 
since
\begin{eqnarray}
U (  \ket{0} \otimes  |  \phi_{n}^{A} \rangle ) 
&\in &
U  \big( E_{1}^{ { Z_{{\bS}}}} \otimes E_{N-2n}^{Z_{\bA}} \big)
\nonumber \\
&\subset& U E_{N+1-2n}^{Z} \nonumber \\
&\subset&  E_{N+1-2n}^{Z} \nonumber \\
&=&   \big( E_{1}^{ { Z_{{\bS}}}} \otimes E_{N-2n }^{ \mathrm{ Z_{\bA}} } \big) \nonumber\\
&&\oplus \big(  E_{-1}^{ { Z_{{\bS}}}} \otimes E_{ N-2(n-1) }^{ \mathrm{ Z_{\bA}} } \big).
\label{U0phi_n=}
\end{eqnarray}
Similarly, the output state $U (  \ket{0} \otimes |  \phi_{0}^{A} \rangle )$ 
is an element of the subspace $ E_{1}^{ { Z_{{\bS}}}} \otimes E_{N }^{ \mathrm{ Z_{\bA}} } $, 
since  
\begin{eqnarray}
U (  \ket{0} \otimes  |  \phi_{0}^{A} \rangle )
&\in &
U  \big( E_{1}^{ { Z_{{\bS}}}} \otimes E_{N}^{Z_{\bA}} \big)
\nonumber \\
&\subset &U E_{N+1}^{Z} \nonumber \\
 &\subset&  E_{N+1}^{Z} \nonumber \\
 &= &   E_{1}^{ { Z_{{\bS}}}} \otimes E_{N }^{ \mathrm{ Z_{\bA}} }. 
\label{U0phi_0=}
\end{eqnarray}
Therefore, by Eqs.~({\ref{U0phi_n=}}) and (\ref{U0phi_0=}),
there exist 
$|  (\phi_{n}^{A})^{0}_{0} \rangle   \in E_{N-2n}^{ Z_{\bA} }$ and 
$|  (\phi_{n-1}^{A})^{0}_{1} \rangle   \in E_{N-2(n-1) }^{ Z_{\bA} }$
 such that
\begin{eqnarray}
U (  \ket{0} \otimes |  \phi_{n}^{A} \rangle )
=    \ket{0}  \otimes |  (\phi_{n}^{A})^{0}_{0} \rangle   
+ \ket{1}  \otimes |  (\phi_{n-1}^{A})^{0}_{1} \rangle,   
\label{U0n}
\end{eqnarray}
where  $|  (\phi_{-1}^{A})^{0}_{1} \rangle = 0 $.
Normalizing Eq.~(\ref{U0n}) gives
\begin{eqnarray}
\| \hspace{0.5mm} |  (\phi_{n}^{A})^{0}_{0} \rangle \|^{2}
+ \| \hspace{0.5mm} |  (\phi_{n-1}^{A})^{0}_{1} \rangle \|^{2} = 1.
\label{phi-01_normalize}
\end{eqnarray}
Similarly, for the output state  $U (  \ket{1} \otimes  |  \phi_{n}^{A} \rangle )$,
there exist 
$|  ({\phi_{n+1}^{A}}) ^{1}_{0} \rangle   \in E_{N-2(n+1) }^{ Z_{\bA} } $ 
and $
| ( {\phi_{n}^{A}} )^{1}_{1} \rangle  \in  E_{ N-2n }^{ Z_{\bA} }$
such that
\begin{eqnarray}
U (  \ket{1} \otimes  |  \phi_{n}^{A} \rangle )
= \ket{0}  \otimes |  ({\phi_{n+1}^{A}}) ^{1}_{0} \rangle   
+ \ket{1}  \otimes | ( {\phi_{n}^{A}} )^{1}_{1} \rangle,
\label{U1n}
\end{eqnarray}
where $
 |  (\phi_{N+1}^{A})^{1}_{0} \rangle = 0
$.  Normalizing Eq.~(\ref{U1n}) gives
\begin{eqnarray}
\| \hspace{0.5mm} |  (\phi_{n+1}^{A})^{1}_{0} \rangle \|^{2}
+ \| \hspace{0.5mm} |  (\phi_{n}^{A})^{1}_{1} \rangle \|^{2} = 1.
\label{phi-10_normalize}
\end{eqnarray}
We can now obtain useful relations
for the output state of $\bS+\bA$ under the conservation law. 
 For the output state $U(\ket{0} \otimes \left| A \right\rangle )$, 
Eqs.~(\ref{A=sum_a_nn})  and (\ref{U0n}) give
\begin{eqnarray}
 U ( \ket{0} \otimes \ket{A} )  
&=&   \ket{0}  \otimes \Big( \sum_{n=0}^{N} a_{n} 
| ( {\phi_{n}^{A}} )^{0}_{0} \rangle   \Big) \nonumber \\ 
& & + \ket{1}  \otimes  \Big( \sum_{n=0}^{N} a_{n}  
 | ( {\phi_{n-1}^{A}} )^{0}_{1} \rangle  \Big).  
\label{U0a_n} 
\end{eqnarray}
Similarly, for the output state $U(\ket{1} \otimes \left| A \right\rangle) $, 
Eqs.~(\ref{A=sum_a_nn}) and (\ref{U1n})  give
\begin{eqnarray}
U( \ket{1} \otimes \ket{A} ) 
&=&   \ket{0}  \otimes \Big( \sum_{n=0}^{N} a_{n} | (\phi_{n+1}^{A})^{1}_{0}  \rangle \Big)  
\nonumber\\
&& + \ket{1}  \otimes  \Big( \sum_{n=0}^{N} a_{n} | (\phi_{n}^{A})^{1}_{1}  \rangle \Big).  
\label{U1a_n}
\end{eqnarray}
Comparing Eq.~(\ref{U0U1}) with Eqs.~(\ref{U0a_n}) and (\ref{U1a_n}), we obtain the following relations:
\begin{eqnarray}
\ket{A^{0}_{0}} &=&  \sum_{n=0}^{N} a_{n} | (\phi_{n}^{A})^{0}_{0}  \rangle,
\nonumber \\\ket{A^{0}_{1}} &=&  \sum_{n=0}^{N} a_{n} | (\phi_{n-1}^{A})^{0}_{1}  \rangle, 
\nonumber \\ 
\ket{A^{1}_{0}} &=&  \sum_{n=0}^{N} a_{n} | (\phi_{n+1}^{A})^{1}_{0}  \rangle, 
\nonumber \\
\ket{A^{1}_{1}} &=&  \sum_{n=0}^{N} a_{n} | (\phi_{n}^{A})^{1}_{1}  \rangle.
\label{correspondence}
\end{eqnarray}

\subsection{Optimization of gate trace distance by ancilla input}

We can now estimate the inner product 
$\langle A^{0}_{1} \ket{A^{1}_{0}}$. 
By Eq.~(\ref{correspondence}),
\begin{eqnarray}
\langle A^{0}_{1} \ket{A^{1}_{0}}  
&=& \sum_{n,n'=0}^{N} {a_{n'}}^{*} a_{n} 
\langle (\phi_{n'-1}^{A})^{0}_{1} |     (\phi_{n+1}^{A})^{1}_{0} \rangle,     
\label{AA=n-1_n+1}
\end{eqnarray}
where the inner product $\langle (\phi_{n'-1}^{A})^{0}_{1} |     (\phi_{n+1}^{A})^{1}_{0} \rangle  $ is given as
\begin{eqnarray}
&& \langle (\phi_{n'-1}^{A})^{0}_{1} |     (\phi_{n+1}^{A})^{1}_{0} \rangle   
\nonumber \\ &&=
\left\{
\begin{array}{l}
0 \hspace{30.8mm}{\mathrm{for}} \hspace{3mm} n'-1\neq n+1,\hspace{5mm}\\
 \langle (\phi_{n+1}^{A})^{0}_{1} |     (\phi_{n+1}^{A})^{1}_{0} \rangle
 \hspace{5mm}{\mathrm{for}} \hspace{3mm} n'-1 = n+1.\hspace{5mm}  
\end{array}
\right. 
\label{i-p_n-1,n+1}
\end{eqnarray}
Therefore,
\begin{eqnarray} 
 \langle A^{0}_{1} \ket{A^{1}_{0}}   &=&
    \sum_{n=0}^{N-2} {a_{n+2}}^{*} a_{n} \langle (\phi_{n+1}^{A})^{0}_{1} |  (\phi_{n+1}^{A})^{1}_{0} \rangle.    
\end{eqnarray}
By the triangle inequality, we have
\begin{eqnarray} | \langle A^{0}_{1} \ket{A^{1}_{0}}  | 
 \leq   
 \sum_{n=0}^{N-2}  |  {a_{n+2}}|\,  | a_{n} |\, 
| \langle (\phi_{n+1}^{A})^{0}_{1} | 
 (\phi_{n+1}^{A})^{1}_{0} \rangle   |. 
\end{eqnarray}
From Eqs.~(\ref{sum_an}), (\ref{phi-01_normalize}), and (\ref{phi-10_normalize}),
the Schwarz inequality gives the relations
\beqa
\sum_{n=0}^{N-2} | {a_{n+2}}| | a_{n} |&\le& 1, \\
| \langle (\phi_{n+1}^{A})^{0}_{1} | 
 (\phi_{n+1}^{A})^{1}_{0} \rangle |
&\leq&  \|  \hspace{0.5mm}   | (\phi_{n+1}^{A})^{0}_{1} \rangle  \| 
\|   | (\phi_{n+1}^{A})^{1}_{0} \rangle  \| 
\leq 1.\nonumber\\
\eeqa
Thus, 
\begin{eqnarray}
\left| \langle A^{0}_{1} \ket{A^{1}_{0}}  \right| 
 & \leq &     \sum_{n=0}^{N-2} | {a_{n+2}}|  | a_{n}| 
\le 1,
\label{|a_n||a_n+2|}
\end{eqnarray}
so that the maximum of $| \langle A^{0}_{1} \ket{A^{1}_{0}}  | $ is at most 
$ \sum_{n=0}^{N-2} | {a_{n+2}} |  | a_{n} |$.  Therefore, 
the minimum of $\frac{1}{2}|1-  \langle A^{0}_{1} \ket{A^{1}_{0}}  | $ in the right-hand side of 
Eq.~(\ref{general_inequality}) is at least $\frac{1}{2}(1-  \sum_{n=0}^{N-2} | {a_{n+2}} |  | a_{n} |)$.  
Since in the above argument the unitary operator $U$ was arbitrary but satisfied the conservation
law, we have 
\begin{eqnarray}
\lefteqn{\min_{U}
\max_{\rho_{{\bS}}}
D \left( \cE_{U,\left| A \right\rangle}( \rho_{{\bS}} ) , \cE_{X_{\bS}}(
\rho_{{\bS}})\right)}\qquad\qquad\qquad\nonumber \\
&\geq&  \frac{1}{2}\Big( 1 -  \sum_{n=0}^{N-2} \left| {a_{n+2}} \right|  \left| a_{n} \right|  \Big),
\label{izon-a_n}
\end{eqnarray}
where $U$ varies over all the unitary operators on $\cH_{\bS}\otimes\cH_{\bA}$ 
satisfying \Eq{c.l.}.
This is a useful inequality that allows us to evaluate a lower bound of 
the quantum NOT gate given the input state of the ancilla system. 
For example, if $a_{n}$ is a constant, such as
\begin{eqnarray}
a_{n}=\frac{1}{\sqrt{N+1}} 
\label{Eq-wei}
\end{eqnarray}
for all $n=0,1,\cdots, N$,
then whatever evolution operator is used,
an error probability $\frac{1}{N+1}$ determined by Eq.~(\ref{izon-a_n}) is unavoidable.

The following questions regarding Eq.~(\ref{izon-a_n}) still remain: 
What is the lower bound over the input states of the ancilla system?
Can we reduce the lower bound to zero by choosing appropriate input states of $\bA$?
In the next section, we will quantitatively show that 
there exists a non-zero lower bound of the error probability 
for any input state of the ancilla system 
and any evolution operator.
In order to obtain the bound, it is necessary to minimize Eq.~(\ref{izon-a_n}) over 
the input states of $\bA$ under condition (\ref{nor-con-a_n}).

\section{Precision limit given the ancilla size \label{accuracy-limit-size}}

We consider the maximization of 
$\sum_{n=0}^{N-2}| {a_{n+2}} | | a_{n} |$ 
over input states of the ancilla system to  minimize 
the right-hand side of Eq.~(\ref{izon-a_n})
under condition (\ref{nor-con-a_n}).
In the first place,
we show that this problem can be reduced to the derivation 
of the maximum eigenvalue of 
a symmetric matrix.
Secondly, we explain how to derive the maximum eigenvalue, 
making use of the recurrence formula of 
Chebyshev polynomials of the second kind. 
We finally describe the lower bound of the quantum NOT gate which depends
only on the size of the ancilla system.  

\subsection{Lower bound and eigenvalue problem}

The summation $\sum_{n=0}^{N-2}| {a_{n+2}} | | a_{n} |$ can be divided into two parts, 
the summation of odd subscripts, such as $| {a_{0}} | | a_{2} |, | {a_{2}} | | a_{4} |$, $\cdots$, and 
that of even subscripts, such as $| {a_{1}} | | a_{3} |, | {a_{3}} | | a_{5} |$, $\cdots$. 
For even $N$,  
\begin{eqnarray}
\lefteqn{\sum_{n=0}^{N-2}  |a_{n+2}|   |a_{n}| }\quad\nonumber \\
&=& \sum_{r=0}^{ \frac{N-4}{2}  }
|a_{2r+1}| |a_{2r+3}|
+  \sum_{r=0}^{ \frac{N-2}{2}   }
|a_{2r}| |a_{2r+2}|, 
\label{sum_even}
\end{eqnarray}
where  
 $N \geq 2$.
For odd $N$,
\begin{eqnarray}
\lefteqn{\sum_{n=0}^{N-2}  |a_{n+2} |  |a_{n}|}\quad \nonumber\\
&=& \sum_{r=0}^{ \frac{N-3}{2}   }
|a_{2r+1}| |a_{2r+3}|
+  \sum_{r=0}^{ \frac{N-3}{2}   }
|a_{2r}| |a_{2r+2}|,
\label{sum_odd}
\end{eqnarray}
where 
 $N \geq 3$.
We now assume that $N$ is even for simplicity; we will comment on the case of odd $N$ later.
To rewrite the summation,  
we define an $(N+1)$-dimensional vector ${\bA}^{\dagger}$ by
\begin{eqnarray}
 {\bA}^{\dagger} 
=\big[\, |a_{1}|,  |a_{3}|,  \ldots, |a_{N-1} |, 
|a_{0}|, |a_{2}|,\ldots, |a_{N}|  \,\big],
\end{eqnarray}
where the odd indexed (resp.\ even indexed) elements are in the first (resp.\ second) 
half elements of the vector, 
and the number of those elements is $\frac{N}{2}$ (resp.\ $\frac{N}{2}+1$).
The summation can then be expressed by a matrix and the vector ${\bA}$ as 
\begin{eqnarray}
\lefteqn{\sum_{n=0}^{N-2}  |a_{n+2}|   |a_{n}| }\nonumber \\
&=& {\bA}^{\dagger}
\left[\begin{array}{cccccc|cccccccc }
0 & 1 & 0&  \cdots & & &0  &\cdots && &0 \\ 
0 & 0 & 1 &  &&&\vdots&&&&\vdots \\
  &     & \ddots & \ddots &&&&& &\\
 &  &  &  0 & 1 &&&&& \\
 &  &  & &0 & &0& \cdots && &0\\
 \hline
0 & \cdots &&&0& & 0 & 1& 0& \cdots & \\
\vdots &&&& \vdots & & 0 & 0 & 1& &        \\
&&&&& &   &   & \ddots & \ddots &  \\
&&&&& &   &   &        &  0 & 1\\
0 &\cdots &&& 0 & &    &     &    &    & 0\\
\end{array}
\right] {\bA}, \qquad
\label{a^T010a}
\end{eqnarray}
where the matrix has four submatrices. 
The upper left (resp.\ lower right) submatrix is the 
$\frac{N}{2} \times \frac{N}{2}$ 
(resp. ($\frac{N}{2} +1) \times (\frac{N}{2}+1 )$ ) matrix with 
all the first subdiagonal 
entries one and all the other entries zero.  
The upper right (resp. lower left) submatrix is the  $\frac{N}{2} \times (\frac{N}{2}+1)$ 
(resp. ($\frac{N}{2} +1) \times \frac{N}{2} $ ) matrix with all the entries zero.
Taking the complex conjugate of both sides of Eq.~(\ref{a^T010a}) gives 
\begin{eqnarray}
\lefteqn{\sum_{n=0}^{N-2}  |a_{n+2}|   |a_{n}|}
\nonumber\\
&=& {\bA}^{\dagger}
\left[
\begin{array}{cccccc|cccccccc }
0 & 0 & \cdots &   &      & &0       &\cdots && &0\\ 
1 & 0 &        &   & & &\vdots  &       && &\vdots  \\
0 & 1 & \ddots &   &       & &        &       && &\\
\vdots &   & \ddots &  0    &0&        &       &&& & \\
 &   &  & 1 &0 & &0&\cdots && &0\\
 \hline
0 &\cdots&&&0& & 0 & 0& \cdots &  & \\
\vdots &&&& \vdots& & 1 & 0 & & & &     \\
&&&&& &  0 &  1 & \ddots &  & \\
&&&&& &   &   &    \ddots    &  0 & \\
0&\cdots && &0& &   &   &        &  1  & 0\\
\end{array}
\right] {\bA}. 
\qquad
\label{a^T000a}
\end{eqnarray}
Therefore, adding Eq.~(\ref{a^T010a}) to Eq.~(\ref{a^T000a}) gives
\begin{eqnarray}
\lefteqn{\sum_{n=0}^{N-2}  |a_{n+2}|   |a_{n}| }
\nonumber \\
&=& {\bA}^{\dagger}
\left[
\begin{array}{cccccc|cccccccc}
0 & \frac{1}{2} & 0&  \cdots & & & 0 & \cdots & & & & 0 \\ 
 \frac{1}{2}  & 0 &  \frac{1}{2}  &  &&& \vdots &&&&& \vdots \\
 0 &     \frac{1}{2}  & 0 &  \ddots  &&&&& &&\\
  &  &   \ddots &  \ddots &  \frac{1}{2} & 0 &&&&& \\
 &  &   & \frac{1}{2} &  0 & \frac{1}{2} & &&&&& \\
 &  &  & 0 & \frac{1}{2}  & 0 &0 & \cdots && && 0 \\
 \hline
0 & \cdots &&&&0 & 0 & \frac{1}{2}& 0& \cdots & \\
\vdots  &&&&& \vdots  & \frac{1}{2} & 0 & \frac{1}{2}& & &\\
&&&&& & 0  &  \frac{1}{2} & 0 & \ddots & \\
&&&&& &   &   &      \ddots   &  \ddots & \frac{1}{2} & 0 \\
&&&&& &   &   &        &   \frac{1}{2} & 0 & \frac{1}{2}\\
0 & \cdots &&&& 0 &   &   &        &  0  & \frac{1}{2} &0\\
\end{array}\right] {\bA}, 
\nonumber \\
\label{1/2-1/2}
\end{eqnarray}
where the upper left and the lower right  submatrices are symmetric with 
all the first subdiagonal and superdiagonal entries $1/2$ and all the other entries 0.
Let  ${\bA}_{\mathrm{odd}}^{\dagger}$ 
and ${\bA}_{\mathrm{even}}^{\mathrm{\dagger}}$ 
be two vectors defined by
\begin{eqnarray}
{\bA}_{\mathrm{odd}}^{\dagger}
&=&  \big[\,  
 |a_{1}|,|a_{3}|, |a_{5}|, \ldots ,  |a_{N-1}| \, \big], 
 \nonumber\\
{\bA}_{\mathrm{even}}^{\mathrm{\dagger}}
&=&  \big[\, 
 |a_{0}|, |a_{2}|,|a_{4}|,  \ldots,  |a_{N}| \,\big], 
\end{eqnarray}
and $S_{l}$ be an $l \times l$ symmetric matrix defined by
\begin{eqnarray}
S_{l}= \left[
\begin{array}{ccccc}
0 & \frac{1}{2} & 0 & 0\\
\frac{1}{2} & 0 &  \frac{1}{2} & 0 \\
0 & \frac{1}{2} & 0 & \ddots \\
0 & 0 & \ddots &\ddots & \frac{1}{2}\\
&&& \frac{1}{2}& 0
\end{array}
\right].
\end{eqnarray}
Then, Eq.~(\ref{1/2-1/2}) can be written as 
\begin{eqnarray}
\sum_{n=0}^{N-2}  |a_{n+2}|   |a_{n}|
&=& {\bA}_{\mathrm{odd}}^{\mathrm{\dagger}}
S_{ \frac{N}{2} } {\bA}_{\mathrm{odd}} 
+ {\bA}_{\mathrm{even}}^{\mathrm{\dagger}}
S_{ \frac{N}{2} +1} {\bA}_{\mathrm{even}} \nonumber \\
&\leq & \|{\bA}_{\mathrm{odd}}\|^{2}\hspace{1mm}s_{ \frac{N}{2} }
+\|{\bA}_{\mathrm{even}}\|^{2}s_{ \frac{N}{2}+1 },
\label{sumbara=a^TT_N-1a} 
\end{eqnarray}
where $s_{l}$ is the maximum eigenvalue of the symmetric matrix $S_{l}$.
Recall that 
$
\| {\bA}_{\mathrm{odd}}\|^{2} + 
\|{\bA}_{\mathrm{even}}\|^{2} 
={\bA}^{\mathrm{\dagger}} \cdot {\bA}  
=1
$, and thus
\begin{eqnarray}
\max_{\sum |a_{n}|^{2}=1} \left[ \sum_{n=0}^{N-2}| {a_{n+2}} | | a_{n} | \right] 
=  \max \big[ s_{ \frac{N}{2}} 
,s_{ \frac{N}{2} +1} \big],
\label{max=tau_N-1}
\end{eqnarray}
where the 
maximization in the right-hand side means selecting 
the larger of $s_{ \frac{N}{2}}$ and 
$s_{\frac{N}{2} +1}$. 

Taking the difference between Eqs. (\ref{sum_even}) and (\ref{sum_odd}) into account,
we apply the same analysis for odd $N$. Then, we have
\begin{eqnarray}
\max_{\sum |a_{n}|^{2}=1} 
\left[ \sum_{n=0}^{N-2}\left| {a_{n+2}} \right| \left| a_{n} \right| \right] 
=   s_{ \frac{N+1}{2}}. 
\label{max=tau_N-1_odd}
\end{eqnarray}
In this way, 
the maximization of the summation 
$ \sum_{n=0}^{N-2}\left| {a_{n+2}} \right| \left| a_{n} \right|$ under condition 
(\ref{nor-con-a_n})
reduces  to the derivation of the maximum eigenvalue of the symmetric matrices  
$S_{\frac{N}{2}}$ and $S_{\frac{N}{2}+1}$.

\subsection{Eigenvalue problem and orthogonal polynomials}

Next we shall determine the maximum eigenvalue, 
as mentioned above, and give the
lower bound  of the quantum NOT gate.
It is well-known that the eigenvalues and the eigenvectors of
the matrix $S_{l}$ are obtained from a
recurrence formula of orthogonal polynomials as follows
\cite{Sze67,Chi78}.
Chebyshev polynomials $W_{l}(x)$  for $l=1,2,\ldots$ 
of the second kind are defined by the relation
\begin{eqnarray}
\label{secondkind}
W_{l}(\cos\theta)=
  \frac{\sin(l+1)\theta}{\sin \theta},
\end{eqnarray}
where $0< \theta < \pi$, and 
 are polynomials of the precise degree $l$, and 
satisfy the recurrence formula  
\begin{eqnarray}
x W_{0}(x)  &=& \frac{1}{2} W_{1}(x),\label{che_l=0} \\
x W_{l} (x) &=& \frac{1}{2}  W_{l+1} (x)  + \frac{1}{2}  W_{l-1}(x),  
\label{Recurrence-Equation}
\end{eqnarray}
where $l\geq 1$.
The roots  $x=x_{l,k}$ of the equation $W_{l} (x)  = 0$ 
is given by
\begin{eqnarray}
x_{l,k}= \cos \frac{k \pi}{l+1} 
\end{eqnarray}
for $k=1,2,\ldots, l$.
Let ${\mathbf{W}}^{\dagger}(x_{l,k})$ be an $l$-dimensional vector defined as 
\begin{eqnarray}
{\mathbf{W}}^{\dagger}(x_{l,k}) 
= \big[
W_{0}(x_{l,k}),  W_{1}(x_{l,k}),  \cdots,   W_{l-1}(x_{l,k})
\big].
\label{def-W^dagger}
\end{eqnarray}
Since $W_{l}(x_{l,k})=0$, Eqs.~(\ref{Recurrence-Equation}) and (\ref{che_l=0}) 
give 
\begin{eqnarray}
S_{l}{\mathbf{W}}(x_{l,k})&=&\left[
\begin{array}{cccccc}
0 & \frac{1}{2} & 0& &  \\ 
\frac{1}{2} & 0 & \frac{1}{2} &  & \\
 0          &  \frac{1}{2} &  0 & \ddots &  \\
 0          &    0      & \ddots & \ddots   & \frac{1}{2}  \\
 &  &  &  \frac{1}{2} &0   \\
\end{array}
\right] 
\left[
\begin{array}{c}
W_{0}(x_{l,k}) \\ W_{1}(x_{l,k}) \\ \vdots \\ \vdots \\ W_{l-1}(x_{l,k})
\end{array}
\right] 
\nonumber \\
&=&
\left[
\begin{array}{c}
\frac{1}{2} W_{1}(x_{l,k}) 
\\ 
\frac{1}{2}W_{0}(x_{l,k}) + \frac{1}{2}W_{2}(x_{l,k})
\\ 
\vdots 
\\ 
\frac{1}{2}W_{j-1}(x_{l,k}) + \frac{1}{2}W_{j+1}(x_{l,k})
\\ 
\vdots
\\
\frac{1}{2}W_{l-2}(x_{l,k}) + \frac{1}{2}W_{l}(x_{l,k})
\end{array}
\right] 
\nonumber \\
&=& x_{l,k}{\mathbf{W}}(x_{l,k}).
\end{eqnarray}
Thus, the vector ${\mathbf{W}}(x_{l,k})$ is an eigenvector of $S_{l}$
 with eigenvalue $x_{l,k}$.
Therefore, the maximum eigenvalue of $S_{l}$ 
 is
\begin{eqnarray}
s_{l} = x_{l,1}=\cos\frac{\pi}{l+1}. \label{tau_N-1=}
\end{eqnarray}
and the corresponding eigenvector is
given by
\begin{eqnarray}
{\mathbf{W}}^{\dagger}(x_{l,1})
&=& 
\displaystyle
\left[ 
\frac{\displaystyle\sin\frac{(j+1)\pi}{l+1}}
{\displaystyle\sin\frac{\pi}{l+1}}\right]_{j=0}^{l-1}.
\end{eqnarray}

\subsection{Derivation of lower bound given the size of ancilla}

We have found the maximum eigenvalue, 
and thus we can now describe a lower bound of the error probability in realizing
 the quantum NOT gate. For even $N$, Eqs.~(\ref{max=tau_N-1}) and (\ref{tau_N-1=})   give    
\begin{eqnarray}
\max_{\sum |a_{n}|^{2}=1}  \sum_{n=0}^{N-2}\left| {a_{n+2}} \right| \left|
a_{n} \right|  = \cos\frac{2\pi}{ N  +4},
\label{max=cos-pi/N}
\end{eqnarray} 
Recall that the minimization of Eq.~(\ref{izon-a_n}) over the input states of $\bA$ is derived from 
the maximization of  $\sum_{n=0}^{N-2}\left| {a_{n+2}} \right| \left| a_{n} \right|$.
Thus, 
\begin{eqnarray}
\lefteqn{\min_{(U,\ket{A})}
\max_{\rho_{{\bS}}}
D (\cE_{U,\ket{A}}( \rho_{{\bS}}),\cE_{X_{\bS}} 
( \rho_{\bS} ) ) }\qquad\nonumber\\
&\geq&  \frac{1}{2}
\Big( 1 -   \cos  \frac{2 \pi}{  N+4  }\Big).  
\label{general_result_even}
\end{eqnarray}
Similarly, for odd $N$	
\begin{eqnarray}
\lefteqn{ \min_{(U,\ket{A})}
\max_{\rho_{{\bS}}}
D (\cE_{U,\left| A \right\rangle}( \rho_{{\bS}}),\cE_{X_{\bS}} 
( \rho_{\bS} ) ) }\qquad
\nonumber \\
&\geq&  \frac{1}{2} \Big(1 -   \cos  \frac{2 \pi}{  N+3   } \Big).  
\label{general_result_odd}
\end{eqnarray}
Here $\cos\frac{2 \pi}{  N+4  } $ is greater than 
$\cos  \frac{2 \pi}{  N+3   } $, and hence we have
finally obtained the lower bound for the error probability of any 
realization of the quantum NOT gate with $N$-qubit control system
under the angular momentum conservation law as
\begin{eqnarray}
\lefteqn{\min_{(U,\ket{A})}
\max_{\rho_{\bS}}
D (\cE_{U,\left| A \right\rangle}( \rho_{\bS} ) , \cE_{X_{\bS}} ( \rho_{{\bS}}) ) }
\qquad\qquad\qquad\nonumber \\
&\geq &
 \frac{1}{2} \Big( 1 -   \cos \frac{2 \pi}{  N+4   } \Big)
\label{general_result}
\end{eqnarray}
for any $N (\geq 2)$.
The bound depends only on the size of the ancilla system: the larger $N$,
the closer to zero is the lower bound.  

\begin{figure}
\centering
\includegraphics[width=8cm,clip]{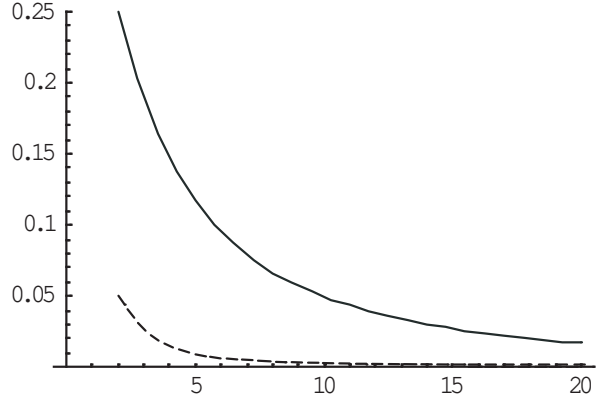}
\caption{   
Plot of the lower bounds as a function of $N$.
The solid line shows the lower bound $\frac{1}{2}(1 -   \cos\frac{2 \pi}{  N+4   }  )$ 
of the quantum NOT gate in Eq.~(\ref{general_result}). 
The dashed line shows the lower bound $\frac{1}{4(N^{2}+1)}$ 
previously obtained for the Hadamard gate in Ref.~\cite{03UPQ}. }
\label{pic}
\end{figure}

According to previous works \cite{02CQC, 03UPQ} 
based on the uncertainty principle, 
it may be expected that the lower bound 
of the quantum NOT gate scales with the inverse of 
$N$ as $\frac{1}{4(N^{2}+1)}\approx\frac{1}{4N^{2}}$.
However, the new bound has the leading order $ \frac{1}{2}( 1 -   \cos \frac{2 \pi}{  N+4   })
\approx\frac{\pi^2}{N^{2}}$, so that the lower bound obtained here is really tighter than that
as depicted by Figure \ref{pic}. 

\subsection{Lower bound: general case}

We have considered the case where the ancilla
state is a pure state.   In the following we shall consider the general case.
Let $(U,\rho_{\bA})$ be a conservative implementation with $N$ qubit ancilla $\bA$.
Then, its purification $(U',\ket{A'})$ is a conservative pure implementation 
with $N+\lceil\log_{2}\rank{\rho_{\bA}}\rceil$ qubit ancilla $\bA'$
such that $\cE_{U,\rho_{\bA}}=\cE_{U',\ket{A'}}$.
Applying \Eq{general_result} to $\cE_{U',\ket{A'}}$, we have
\beqa
\lefteqn{
\max_{\rho_{\bS}}D(\cE_{U,\rho_{\bA}}(\rho_\bS),
\cE_{X_{\bS}}(\rho_\bS)))}\quad\nonumber\\
&\ge& 
 \frac{1}{2}\Big(  1-  \cos  \frac{2\pi}{N+\log_{2}\rank{\rho_{\bA}}
+4}\Big),
\eeqa
and from $N+\log_{2}\rank{\rho_{\bA}}\le 2N$, and we conclude 
\beqa
\lefteqn{
\min_{(U,\rho_{\bA})}\max_{\rho_{\bS}}
D(\cE_{U,\rho_{\bA}}(\rho_\bS),\cE_{X_{\bS}}(\rho_\bS))}\qquad\qquad\qquad\qquad\nonumber\\
&\ge&
 \frac{1}{2}\Big(   1-  \cos \frac{\pi}{N+2}\Big),
\eeqa
where $(U,\rho_{\bA})$ varies over all the conservative implementations 
with $N$ qubit ancilla.

\section{Lower bounds for classically complete implementations 
and their attainability \label{a-noteworthy}}

In the preceding section, we have shown that a general lower bound for
the error probability in realizing the quantum NOT gate is given by the 
$1-\cos(1/N)$ scale for the ancilla size $N$, instead of $1/N^{2}$
scaling already known for some other gates.  
Since $2[1-\cos(1/N)]= 1/N^{2}-1/(12N^{4})+\cdots$,
the new scale has the same leading order as $1/N^{2}$ up to constant,
but it is natural to ask if the higher order terms are really meaningful.
Here, we shall answer this question, so that the 
$1-\cos(1/N)$ scale is the best result.
To show this, we shall show the attainability of
a lower bound with  the $1-\cos(1/N)$ scale
for classically complete conservative pure implementations.
Thus, a classically complete conservative implementation exists even with 
only 2 qubit ancilla, whereas the substantial error occurs when the input 
state is a superposition of computational basis states.
This result also shows that the general lower bound for conservative 
implementations with $N$ qubit ancilla can be reached by a
classically complete conservative pure implementations with $2N$ qubit 
ancilla.  

\subsection{Classically complete pure implementations}

Let $(U',\ket{A'})$ be a classically complete conservative pure implementation.
Then,  we have the following relations 
\begin{eqnarray}
U' \left( \ket{0} \otimes \ket{A'} \right) &=&
\ket{1} \otimes | {A'}^{0}_{1} \rangle, \nonumber \\
U' (\ket{1} \otimes \ket{A'}) &=&
\ket{0} \otimes | {A'}^{1}_{0} \rangle,  
\label{special-U'-Ancilla}
\end{eqnarray}
where $ | {A'}^{0}_{1} \rangle$ and 
$| {A'}^{1}_{0} \rangle \in \cH_{\bA}$. 

First, we discuss the constraint on 
the input state $\ket{A'}$ of $\bA$ imposed by the above relations.
To illustrate this, we describe $\ket{A'}$ as 
\begin{eqnarray}
\ket{A'} = \sum_{n=0}^{N}a'_{n} \ket{\phi_{n}^{A'}},
\label{|A'>:=}
\end{eqnarray}
where $\ket{\phi_{n}^{A'}}$  are 
 normalized vectors in the eigenspaces $E_{N-2n}^{Z_{\bA}}$
for all $n=0,1,\cdots, N$,
and we have $\sum_{n=0}^{N}|a'_{n}|^{2}=1$.
Suppose that the input state of $\bS$ is $\ket{0}$.  
Recalling that relation  (\ref{U0phi_0=}) holds by the conservation law,
the output state corresponding to the input state
$\ket{0} \otimes | \phi_{0}^{A'} \rangle$ can be written as
\begin{eqnarray}
U'( \ket{0} \otimes | \phi_{0}^{A'} \rangle)
& = & e^{i \phi'} \ket{0} \otimes | \phi_{0}^{A'} \rangle,
\end{eqnarray}
where $e^{i \phi'}$ is a phase factor. 
Thus the output state corresponding to the input state $\ket{0} \otimes \ket{A'} $
can be expressed as 
\begin{eqnarray}
\lefteqn{U' (\ket{0} \otimes \ket{A'})}\quad\nonumber\\
& =&  a'_{0}e^{i \phi'} \ket{0} \otimes\ket{\phi_{0}^{A'}}+
\sum_{n=1}^{N}a'_{n}U'(\ket{0} \otimes \ket{\phi_{n}^{A'}}) .
\end{eqnarray}
Comparing with Eq.~(\ref{special-U'-Ancilla}),
$a'_{0}$ must be zero.
Similarly,
$a'_{N}$ must be zero, considering the input state $\ket{1}$.  	

We now describe the output state in $\bS$ from $(U',\ket{A'})$ 
for any pure input state $\ket{\psi}$.
 This is given by the partial trace of  the output state in $\bS+\bA$ with respect to $\bA$:
\begin{eqnarray}
\lefteqn{ \cE_{U',\ket{A'}}( \ket{\psi})}
\nonumber \\ 
&=& 
 {\mathrm{Tr}}_{\bA} 
 \left[ U' ( \ket{\psi} \otimes \ket{A'} )\left( 
\langle \psi | \otimes \langle A' | 
\right) {U'}^{\dagger}
\right]
\nonumber\\
&=& |\beta |^{2} \ket{0}\bra{0}
 +\alpha^{*}\beta  \langle {A'}^{0}_{1} \big| {A'}^{1}_{0} \rangle \ket{0}\bra{1} 
+\alpha \beta^{*} \langle {A'}^{1}_{0} \big| {A'}^{0}_{1} \rangle \ket{1}\bra{0}
\nonumber \\
 & &+|\alpha|^{2} \ket{1}\bra{1}. 
\label{rhoU'ij}
\end{eqnarray}
Here, we use abbreviation such as $\cE( \ket{\psi}):=\cE( \ket{\psi}\bra{\psi} )$ for any
operation $\cE$.
The trace distance between the ideal quantum NOT operation (\ref{rhoXij}) 
and $\cE_{U',\ket{A'}}( \left| \psi \right\rangle )$ is then 
\begin{eqnarray}
D(\cE_{X_{\bS}}( \left| \psi \right\rangle) , \cE_{U',\ket{A'} }( \left| \psi \right\rangle ) ) 
= \left| \alpha^{*}\beta \right|  
|   1- \langle {A'}^{0}_{1} \big| {A'}^{1}_{0} \rangle   |.\quad
\label{D(X,U')=|1-A|}
\end{eqnarray}
Thus, the derivation of the lower bound for the gate implementation $(U',\ket{A'})$ 
can be reduced to estimating the maximum value  of $\langle {A'}^{0}_{1} | {A'}^{1}_{0} \rangle$, 
which is very similar to the general analysis of Sec.~\ref{lowerbound-conservation}. 
However, this case  differs 
from the general analysis in that  $a_{0}=a_{N}=0$. 	
Taking this condition into account,  
$| {A'}^{0}_{1} \rangle$ and $| {A'}^{1}_{0} \rangle$ can be 
written  
as 
\begin{eqnarray}
| {A'}^{0}_{1} \rangle &=&  \sum_{n=1}^{N-1} a'_{n} | (\phi_{n-1}^{A'})^{0}_{1}  \rangle, 
\nonumber \\ 
| {A'}^{1}_{0} \rangle &=&  \sum_{n=1}^{N-1} a'_{n} | (\phi_{n+1}^{A'})^{1}_{0}  \rangle, 
\label{correspondence'}
\end{eqnarray}
where $| (\phi_{n-1}^{A'})^{0}_{1}  \rangle$ and $| (\phi_{n+1}^{A'})^{1}_{0}  \rangle $ 
are normalized vectors in the eigenspaces $E_{N-2(n-1)}^{Z_{\bA}} $ 
and  $E_{N-2(n+1)}^{Z_{\bA}}$, respectively.
Thus,
\begin{eqnarray}
| \langle {A'}^{0}_{1} | {A'}^{1}_{0} \rangle  | 
 & \leq &     \sum_{n=1}^{N-3} | {a'_{n+2}} |  | a'_{n} |,
\label{|a_n'||a_n+2'|}
\end{eqnarray}
and therefore,
\begin{eqnarray}
\lefteqn{\min_{U'}
\max_{\rho_{{\bS}}}
D (  \cE_{U',\ket{A'}}( \rho_{{\bS}}),\cE_{X_{\bS}} ( \rho_{\bS} )) }\qquad\qquad\qquad
\nonumber\\
&\geq&  \frac{1}{2}
\Big( 1 -    \sum_{n=1}^{N-3} | {a'_{n+2}} |  | a'_{n} | \Big).  
\label{special_result-pre}  
\end{eqnarray}
Since the discussion in Sec.~\ref{accuracy-limit-size} can be applied 
to minimizing Eq.~(\ref{special_result-pre}) over the input states of $\bA$, 
we see that for even $N$
\begin{eqnarray}
\lefteqn{
\min_{(U',\ket{A'})}
\max_{\rho_{{\bS}}}
D (\cE_{U',\ket{A'}}( \rho_{{\bS}}),\cE_{X_{\bS}} ( \rho_{\bS} ) ) }
\qquad\qquad\qquad
\nonumber \\
&\geq&  \frac{1}{2}
\Big( 1 -   \cos\frac{2\pi}{  N  +2} \Big).  
\label{special_result}  
\end{eqnarray}
This lower bound is slightly larger than the one 
for the general case; the difference 
comes close to zero for large $N$ of the ancilla system.
We shall comment on the odd $N$ case later.

\subsection{Attainability of the lower bound for classically complete pure implementations}

Next we show that there exists a classically complete implementation
$(U',\ket{A'})$ which attains the lower bound 
$\frac{1}{2}(1-   \cos \frac{2\pi}{  N  +2} )$.
We begin by describing the input state $|  \tilde{A}  \rangle $ as follows.
Let $| (e_{n})^{i} \rangle$ be fixed orthonormal bases in eigenspace 
$E_{N-2n}^{Z_{\bA}}$ as 
\begin{eqnarray}
| (e_{n})^{1} \rangle, | (e_{n})^{2} \rangle,\cdots,
| (e_{n})^{k} \rangle,\cdots, | (e_{n})^{d_{n}} \rangle, 
\end{eqnarray}
for $n=0,1, \cdots, N$,
where $d_{n}=\frac{N!}{n!(N-n)!}$. In addition, 
$\tilde{{\bA}}_{\mathrm{odd}}^{\dagger}$  and 
$\tilde{{\bA}}_{\mathrm{even}}^{\mathrm{\dagger}}$
are two vectors:  
\begin{eqnarray}
\tilde{{\bA}}_{\mathrm{odd}}^{\dagger}
&=&  \big[\,  \tilde{a}_{1},  \tilde{a}_{3},\tilde{a}_{5},  \cdots , \tilde{a}_{N-1}\,\big], 
 \nonumber\\
\tilde{{\bA}}_{\mathrm{even}}^{\mathrm{\dagger}}
&=&  \big[ \,
 \tilde{a}_{2},  \tilde{a}_{4},\tilde{a}_{6}, \cdots ,  \tilde{a}_{N-2}\,\big]. 
\end{eqnarray}
where $\tilde{{\bA}}_{\mathrm{odd}}^{\dagger}$ 
(resp.\ $\tilde{{\bA}}_{\mathrm{even}}^{\dagger}$) is a $\frac{N}{2}$ 
(resp.\ $\frac{N}{2}-1$) dimensional 
vector whose entries are indexed by odd (resp.\ even) numbers.
We assume that these vectors satisfy 
\begin{eqnarray}
\tilde{\bA}_{\mathrm{odd}} 
&=& \frac{1}{C_{ \frac{N}{2} }}\mathbf{W} (x_{ \frac{N}{2} ,1}),\nonumber \\
\tilde{\bA}_{\mathrm{even}}
&=& 0,
\label{distribution_ancilla_input}
\end{eqnarray}
where  
$C_{ \frac{N}{2}  }
= [ 
{\mathbf{W}} (x_{ \frac{N}{2} ,1} )^{\dagger }
{\mathbf{W}} (x_{ \frac{N}{2} ,1} ) ]^{\frac{1}{2}}$. It follows that 
$\|{\tilde{{\bA}}}_{\mathrm{odd}}\|^{2} =1$ by 
normalization.
We assume that the input state $|  \tilde{A}  \rangle$ is given by
\begin{eqnarray}
|  \tilde{A}  \rangle 
= \sum_{n=1}^{N-1} \tilde{a}_{n} | (e_{n})^{1} \rangle.
\end{eqnarray}
Recall that 
${\mathbf{W}} (x_{ \frac{N}{2}, 1} ) $ is an eigenvector with the maximum eigenvalue 
of $S_{ \frac{N}{2} }$. 
Then the coefficients  $\tilde{a}_{n}$ satisfy the following equation:
\begin{eqnarray}
\sum_{n=1}^{N-3}  \tilde{a}_{n+2} \hspace{1mm}  \tilde{a}_{n}
&=& \tilde{{\bA}}_{\mathrm{odd}}^{\mathrm{\dagger}}
S_{ \frac{N}{2} } \tilde{{\bA}}_{\mathrm{odd}} 
 \nonumber \\
&=&\frac{1 }{ C_{ \frac{N}{2} }^{2}} 
{\mathbf{W}} (x_{ \frac{N}{2} ,1} )^{\dagger }
S_{ \frac{N}{2} } 
{\mathbf{W}} (x_{ \frac{N}{2}, 1} )
\nonumber \\ 
&=&   s_{ \frac{N}{2}}
\nonumber \\
&=& \cos\frac{2\pi}{ N +2}.
\label{a_{n}_attain_cos}
\end{eqnarray}

Constructing the evolution operator $\tilde{U}$ can be accomplished 
by determining the transformation for all orthonormal bases.
We require that $\tilde{U}$ satisfy the following conditions. 
For $n =1, 2,  \cdots, N$,  
\begin{eqnarray}
\tilde{U} \left( \ket{0} \otimes | (e_{n})^{1} \rangle \right)
&=&  \ket{1} \otimes | (e_{n-1})^{1} \rangle,  
\nonumber \\
\tilde{U} \left( \ket{1} \otimes | (e_{n-1})^{1} \rangle \right)
&=& \ket{0} \otimes | (e_{n})^{1} \rangle, 
\label{deftildeUn}
\end{eqnarray}
and for all bases except those that appear in Eq.~(\ref{deftildeUn}),
\begin{eqnarray}\label{xxx}
\tilde{U}\left( \ket{0} \otimes | (e_{n})^{i} \rangle \right)
&=&\ket{0} \otimes | (e_{n})^{i} \rangle,  
\nonumber \\
\tilde{U} \left( \ket{1} \otimes | (e_{n})^{i} \rangle \right)
&=& \ket{1} \otimes | (e_{n})^{i} \rangle.
\end{eqnarray}
These requirements determine one-to-one mapping on the orthonormal basis,
$\{\ket{0}\otimes| (e_{n})^{i} \rangle,\ket{1}\otimes| (e_{n})^{i} \rangle\}$,
and hence there uniquely exists a unitary operator $\tilde{U}$ fulfilling the
above requirements.
Note also that $\tilde{U}$ 
satisfies the conservation law (\ref{c.l.}), since from 
Eqs.~(\ref{deftildeUn}) and (\ref{xxx}) we have the relations
$ UE_{\lambda}^{Z}  \subset E_{\lambda}^{Z}$
for all $\lambda$,
which are equivalent to the conservation law, as seen in Eq.~(\ref{[U,Z]=0arrowUEsubsetE}). 

We now describe the output 
state of $(\tilde{U}, |  \tilde{A}  \rangle)$ and the trace distance 
between the ideal output state and that of $(\tilde{U}, |  \tilde{A}  \rangle) $.
The output states for $\ket{0}$ and $\ket{1}$ can be generally written as 
\begin{eqnarray}
\tilde{U} ( \ket{0} \otimes | \tilde{A} \rangle )
&=& \ket{0}  \otimes | \tilde{A}^{0}_{0} \rangle  
+ \ket{1}  \otimes | \tilde{A}^{0}_{1} \rangle,  \nonumber\\
\tilde{U} ( \ket{1} \otimes | \tilde{A} \rangle )
&=& \ket{0}  \otimes | \tilde{A}^{1}_{0} \rangle  
+ \ket{1}  \otimes | \tilde{A}^{1}_{1} \rangle,  
\label{tildeU0U1}
\end{eqnarray}
respectively, where $| \tilde{A}^{i}_{j} \rangle \in \cH_{\bA}$ with  $i,j=0,1$.  
 On the other hand,  by the definitions of $\tilde{U} $ and $ |  \tilde{A}  \rangle $, we have
\begin{eqnarray}
\tilde{U}  ( |0 \rangle \otimes |  \tilde{A}  \rangle )
&=& \tilde{U}  \Big( |0 \rangle \otimes  \sum_{n=1}^{N-1} \tilde{a}_{n} | (e_{n})^{1} \rangle \Big)
\nonumber \\
&=& |1 \rangle \otimes \Big( \sum_{n=1}^{N-1} \tilde{a}_{n}  | (e_{n-1})^{1} \rangle  \Big),
\nonumber \\
\tilde{U}  ( |1 \rangle \otimes |  \tilde{A}  \rangle )
&=& \tilde{U}  \Big( \ket{1} \otimes  \sum_{n=1}^{N-1} \tilde{a}_{n} | (e_{n})^{1} \rangle \Big)
\nonumber \\
&=&  |0 \rangle \otimes \Big( \sum_{n=1}^{N-1} \tilde{a}_{n}   | (e_{n+1})^{1} \rangle  \Big).
\label{tilde_0_sum}
\end{eqnarray}
Thus  we have the following relations:
\begin{eqnarray}
| \tilde{A}^{0}_{0} \rangle &=&  0,
\nonumber \\
| \tilde{A}^{0}_{1} \rangle &=& 
\sum_{n=1}^{N-1}  \tilde{a}_{n}  | (e_{n-1})^{1} \rangle,   \nonumber \\
| \tilde{A}^{1}_{0} \rangle &=& 
 \sum_{n=1}^{N-1} \tilde{a}_{n} | (e_{n+1})^{1} \rangle, 
\nonumber \\
| \tilde{A}^{1}_{1} \rangle &=& 0.
\label{tilde_correspondence}
\end{eqnarray}
Let $\cE_{\tilde{U}, | \tilde{A} \rangle}( \ket{\psi} )$ be the 
output state of $\bS$ from $(\tilde{U}, | \tilde{A} \rangle) $.
The trace distance between $\cE_{X_{{\bS}}}( \ket{\psi} )$ and
 $\cE_{\tilde{U}, | \tilde{A} \rangle}( \ket{\psi} )$
can be expressed in the same way as for Eq.~(\ref{D()_expre_as_epsilon}) so that we have
\begin{eqnarray}
\lefteqn{D(\cE_{\tilde{U},| \tilde{A} \rangle}( \left| \psi \right\rangle ),
\cE_{X_{{\bS}}}( \left| \psi \right\rangle) )}\quad
\nonumber \\
&=& \Big\{ \big| \alpha^{*}\beta \left( 1- \langle \tilde{A}^{0}_{1} | \tilde{A}^{1}_{0} \rangle \right)
+ \alpha \beta^{*}  \langle \tilde{A}^{1}_{1} | \tilde{A}^{0}_{0} \rangle 
\nonumber \\
& & 
- |\alpha |^{2} \langle \tilde{A}^{0}_{1} | \tilde{A}^{0}_{0} \rangle
 -|\beta |^{2}  \langle \tilde{A}^{1}_{1} | \tilde{A}^{1}_{0} \rangle
\big|^{2}   \nonumber \\ 
& & 
+  \big[ \left(  -|\alpha|^{2} \tilde{\epsilon}_{0} + |\beta|^{2} \tilde{\epsilon}_{1} \right) 
\nonumber \\
& & 
- 2 \mathrm{Re}\big(
\alpha^{*}\beta   \langle \tilde{A}^{0}_{0} | \tilde{A}^{1}_{0} \rangle   
  \big) \big]^{2}  \Big\}^{\frac{1}{2}},
\end{eqnarray}
where
$\|  | \tilde{A}^{0}_{0} \rangle  \|^{2}  = \tilde{\epsilon}_{0}$,
$\|  | \tilde{A}^{1}_{1} \rangle  \|^{2}  = \tilde{\epsilon}_{1} $. However, in this case,
 $\tilde{\epsilon}_{0} = \tilde{\epsilon}_{1} = 0$ from Eq.~(\ref{tilde_correspondence}), and therefore 
\begin{eqnarray}
D( \cE_{\tilde{U},| \tilde{A} \rangle}( \left| \psi \right\rangle ),\cE_{X_{{\bS}}}( \left| \psi \right\rangle) )
&=& \big| \alpha^{*} \beta  ( 1- \langle \tilde{A}^{0}_{1} | \tilde{A}^{1}_{0}  \rangle )
\big|. \nonumber 
\end{eqnarray}
Recall that $| (e_{n})^{1} \rangle$ are orthonormal bases.
Then, Eq.~(\ref{a_{n}_attain_cos}) gives
\begin{eqnarray}
\langle \tilde{A}^{0}_{1} | \tilde{A}^{1}_{0} \rangle 
&=&   \sum_{n, n'=1}^{N-1}  \tilde{a}_{n} \tilde{a}_{n'}  \langle (e_{n-1})^{1} | (e_{n'+1})^{1} \rangle  
\nonumber \\
&=&   \sum_{n' =1}^{N-3}  \tilde{a}_{n'+2} \tilde{a}_{n'}   
\nonumber \\
&=&   \cos \frac{2\pi}{ N  +2}.
\end{eqnarray}
Thus, 
\begin{eqnarray}
D( \cE_{\tilde{U},| \tilde{A} \rangle}( \left| \psi \right\rangle ),
\cE_{X_{{\bS}}}( \left| \psi \right\rangle)  )
&=& \Big| \alpha^{*} \beta  \Big( 1- \cos  \frac{2\pi}{ N  +2 } \Big) 
\Big|. \nonumber 
\end{eqnarray}
Since the right-hand side is maximized where $| \alpha^{*} \beta| = \frac{1}{2}$,
we have
\begin{eqnarray}
\lefteqn{
\max_{ \ket{\psi}}
D(  \cE_{\tilde{U},| \tilde{A} \rangle}( \ket{\psi} ) ,\cE_{X_{{\bS}}}( \ket{\psi}))}\quad
 \nonumber \\
&=&  \frac{1}{2}\Big(  1-  \cos \frac{2\pi}{N +2}\Big).
\end{eqnarray}

\begin{figure}
\centering
\includegraphics[width=8cm,clip]{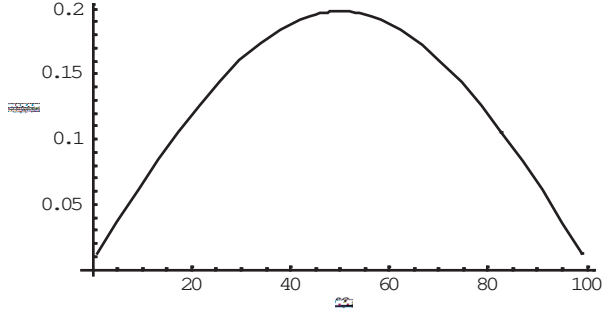}
\caption{Distribution of 
$|a_{n}|$ with odd subscripts for $N=100$ which gives the lower bound in Eq.~(\ref{special_result}).
This figure shows 
$ 
\frac{1}{C_{ \frac{N}{2} }}W_{\frac{n-1}{2}} (x_{ \frac{N}{2} ,1} )$ as a function of odd $n$.}
\label{fig2}
\end{figure}

That is, the model $(\tilde{U}, |  \tilde{A}  \rangle)$ attains the lower bound in
 Eq.~(\ref{special_result}).
Notice that  our model $(\tilde{U}, |  \tilde{A}  \rangle)$ has
a distribution of  $|a_{n}|$, as given by Eq.~(\ref{distribution_ancilla_input}).
Figure \ref{fig2} describes the distribution for $N=100$. 
From a qualitative point of view, 
in order to reduce the lower bound of the quantum NOT gate, 
an input state of the ancilla system should be prepared
which has a sufficiently thick distribution in the neighborhood of eigenvalue 0,
rather than a constant distribution,  such as that given by Eq.~(\ref{Eq-wei}).

For odd $N$,  the lower bound can be given 
by setting the input state and the evolution operator 
as those analogous to the case of  even $N$.
The bound is
 $ \frac{1}{2} ( 1-  \cos \frac{2\pi}{ N +1} )$.
The attainability of this bound is also  proved by the analogous argument.

Thus, we have shown that 
\beqa
\lefteqn{
\min_{(U,\ket{A})}\max_{\rho_{\bS}}
D (\cE_{U,\left| A \right\rangle}( \rho_{\bS}),\cE_{X_{\bS}} ( \rho_{\bS} )  ) }\quad\nonumber\\
&=& \frac{1}{2}\Big(   1-  \cos \frac{2\pi}{N +2}\Big),
\eeqa
if $N$ is even and 
 \beqa
\lefteqn{
\min_{(U,\ket{A})}\max_{\rho_{\bS}}
D (\cE_{U,\left| A \right\rangle} ( \rho_{\bS} ) , \cE_{X_{\bS}}( \rho_{{\bS}}) ) }\quad\nonumber\\
&=&
\frac{1}{2}\Big(   1-  \cos \frac{2\pi}{N +1}\Big)  
\eeqa
if $N$ is odd, where $(U,\ket{A})$ varies over all the classical complete pure implementation with
$N$ qubit ancilla.

For arbitrary $N$, we conclude as a common lower bound
 \beqa
\lefteqn{
\min_{(U,\ket{A})}\max_{\rho_{\bS}}
D (\cE_{U,\left| A \right\rangle}( \rho_{{\bS}}),\cE_{X_{\bS}} ( \rho_{\bS} )  ) }\quad\nonumber\\
&\ge&
\frac{1}{2}\Big( 1-  \cos \frac{2\pi}{N +2}\Big),
\eeqa
where $(U,\ket{A})$ varies over all the classical complete pure implementation with
$N$ qubit ancilla.

We have considered the case where the ancilla
state is a pure state.  
The lower bound for the general case is obtained by the previously developed
purification argument, and we conclude the following relations.
We have
\beqa
\lefteqn{
\max_{\rho_{\bS}}
D(\cE_{U,\rho_{\bA}}(\rho_\bS),\cE_{X_{\bS}}(\rho_\bS))}\quad\nonumber\\
&\ge&
 \frac{1}{2}\Big(  1-  \cos  \frac{2\pi}{N+\log_{2}\rank{\rho_{\bA}}+2}\Big),
\eeqa
for any classically complete implementation $(U,\rho_{\bA})$
with $N$ qubit ancilla, and 
\beqa
\lefteqn{
\min_{(U,\rho_{\bA})}\max_{\rho_{\bS}}
D(\cE_{U,\rho_{\bA}}(\rho_\bS),\cE_{X_{\bS}}(\rho_\bS))}\qquad\qquad\qquad\nonumber\\
&\ge&
 \frac{1}{2}\Big(1-  \cos \frac{\pi}{N+1}\Big),
\eeqa
where $(U,\rho_{\bA})$ varies over all the classically complete implementation 
with $N$ qubit ancilla.

\section{Concluding remarks}

In this paper, we have studied the precision limit of the quantum NOT gate 
or the bit flip gate,  one of the most basic gates in quantum computation,
represented on the single-spin computational qubit by considering
the angular momentum conservation law obeyed by the interaction between
the computational qubit and the control system supposed to comprise  
many qubits.  
Actually, we have considered the effect of the angular 
momentum conservation law only in the direction same as the computational
basis, usually set as the $z$ direction.  Then, the conserved quantity and
the computational basis are represented by the Pauli $Z$ operator, whereas 
the quantum NOT gate is represented by the Pauli $X$ operator.  
Thus, it is expected that this non-commutativity leads to a precision limit 
of the gate operation.

In the previous method which was used for other gates \cite{02CQC,03UPQ},
one finds a way in which the gate under consideration is used as a 
component of a measuring apparatus, applies the quantitative 
generalization of the Wigner-Araki-Yanase (WAY) theorem to this 
measuring apparatus, and obtains the lower bound of error probability. 
For the Hadamard gate, one finds that 
it is used to convert the $Z$ measurement to the $X$ measurement,
and that $Z$ measurement can be done without error under
the conservation law of the $z$ component. Then, one can
conclude that the inevitable error of the $X$ measurement,
calculated from the quantitative version of the WAY theorem, 
is yielded from the converter using the Hadamard gate. 
This and similar arguments cannot be applied
to the quantum NOT gate, since the quantum NOT gate
does not convert the direction of measurement, 
but simply flips the measured bit.

In this paper, we have developed a new method for obtaining
the inevitable error probability by evaluating the maximum 
trace distance between the output from
the gate realization and the output from the ideal gate. 
The previous method naturally leads to a
lower bound for the infidelity (one minus the squared fidelity).  
Since the infidelity is dominated by
the trace distance, the new method gives a tighter lower bound for the error 
probability.
 
The new method is based on a straightforward evaluation 
of the trace distance of two output states,
and enables us to find the precision limit Eq.~(\ref{izon-a_n}), explicitly
described by the input state of the ancilla system.  It is thus
possible to obtain information on how much an ancilla
input has an inherent error probability in itself. The
correspondence between the two methods is not easy to
elicit, but it is an interesting problem for future studies that
would lead to a deeper understanding of precision limits
to quantum control systems.

We have also obtained the lower bound (\ref{general_result}) expressed
by the size of the ancilla system, by minimizing Eq.~(\ref{izon-a_n}) 
over the input states of $\bA$, using Chebyshev polynomials of the second kind. 
The lower bound
is much tighter than the scaling expected from the previous result based
on the WAY theorem. 
Since the quantitative generalization of the WAY theorem has a close
relation to the universal uncertainty principle for measurement and disturbance
\cite{03UVR,03UPQ},  
the previous lower bound for pure conservative implementations
is based on the variance of the ancilla state, 
and scales as $\frac{1}{4N^{2}+4}\approx\frac{1}{4N^{2}}$, 
whereas the new method revealed 
the lower bound $\frac{1}{2}(1-\cos\frac{2\pi}{N+4})
\approx \frac{\pi^{2}}{N^{2}}$ as a tighter bound.
The higer order terms in $\frac{1}{2}(1-\cos\frac{2\pi}{N+4})$ is
considered to be meaningful, since the lower bound $\frac{1}{2}(1-\cos\frac{2\pi}{N+2})$
is attained among classically complete pure conservative implementations.
Interestingly, the attainability result shows
that the best ancilla states to attain the lower bound are
not maximum variance states, nor uniformly distributed
states, but those states with the distribution determined
by the recurrence relation characterized by Chebyshev
polynomials.

Although our study has assumed that the ancilla system consists
 of $N$ qubits for comparison with
the previous research, the present method is not restricted to
this particular control system, and it can be readily applied
to other control systems, such as atom-field systems,
where the present method would lead to a lower bound that
scales as the inverse of the photon number \cite{05CQL}. 
Our method will be also expected to contribute to the problem 
of programmable quantum processors
\cite{NC97,VC00,HZB06} and related subjects 
\cite{DP05,DP05b,DP05c} in future investigations.

\begin{acknowledgments}
The authors thank Hajime Tanaka, Gen Kimura, and Julio Gea-Banacloche 
for useful discussions and suggestions.  This research was
partially supported by the SCOPE project of the MIC, the  Grant-in-Aid
for Scientific Research (B)17340021 of the JSPS, and the CREST project of the JST.
\end{acknowledgments}


\begin{thebibliography}{38}
\expandafter\ifx\csname natexlab\endcsname\relax\def\natexlab#1{#1}\fi
\expandafter\ifx\csname bibnamefont\endcsname\relax
  \def\bibnamefont#1{#1}\fi
\expandafter\ifx\csname bibfnamefont\endcsname\relax
  \def\bibfnamefont#1{#1}\fi
\expandafter\ifx\csname citenamefont\endcsname\relax
  \def\citenamefont#1{#1}\fi
\expandafter\ifx\csname url\endcsname\relax
  \def\url#1{\texttt{#1}}\fi
\expandafter\ifx\csname urlprefix\endcsname\relax\def\urlprefix{URL }\fi
\providecommand{\bibinfo}[2]{#2}
\providecommand{\eprint}[2][]{\url{#2}}

\bibitem[{\citenamefont{Shor}(1994)}]{Sho94}
\bibinfo{author}{\bibfnamefont{P.~W.} \bibnamefont{Shor}}, in
  \emph{\bibinfo{booktitle}{Proceedings of the 35th Annual Symposium on
  Foundations of Computer Science}}, edited by
  \bibinfo{editor}{\bibfnamefont{G.}~\bibnamefont{Goldwasser}}
  (\bibinfo{publisher}{IEEE Computer Society Press}, \bibinfo{address}{Los
  Alamitos, CA}, \bibinfo{year}{1994}), pp. \bibinfo{pages}{124--134}.

\bibitem[{\citenamefont{Unruh}(1995)}]{Unr95}
\bibinfo{author}{\bibfnamefont{W.~G.} \bibnamefont{Unruh}},
  \bibinfo{journal}{Phys.\ Rev.\ A} \textbf{\bibinfo{volume}{{51}}},
  \bibinfo{pages}{992} (\bibinfo{year}{1995}).

\bibitem[{\citenamefont{Palma et~al.}(1996)\citenamefont{Palma, Suominen, and
  Ekert}}]{PSE96}
\bibinfo{author}{\bibfnamefont{G.~M.} \bibnamefont{Palma}},
  \bibinfo{author}{\bibfnamefont{K.~A.} \bibnamefont{Suominen}},
  \bibnamefont{and} \bibinfo{author}{\bibfnamefont{A.~K.} \bibnamefont{Ekert}},
  \bibinfo{journal}{Proc.\ R. Soc.\ Lond.\ A} \textbf{\bibinfo{volume}{{452}}},
  \bibinfo{pages}{567} (\bibinfo{year}{1996}).

\bibitem[{\citenamefont{Haroche and Raimond}(1996)}]{HR96}
\bibinfo{author}{\bibfnamefont{S.}~\bibnamefont{Haroche}} \bibnamefont{and}
  \bibinfo{author}{\bibfnamefont{J.-M.} \bibnamefont{Raimond}},
  \bibinfo{journal}{Physics Today} \textbf{\bibinfo{volume}{49}},
  \bibinfo{pages}{no. 8, p. 51} (\bibinfo{year}{1996}).

\bibitem[{\citenamefont{Shor}(1995)}]{Sho95b}
\bibinfo{author}{\bibfnamefont{P.~W.} \bibnamefont{Shor}},
  \bibinfo{journal}{Phys.\ Rev.\ A} \textbf{\bibinfo{volume}{{52}}},
  \bibinfo{pages}{R2493} (\bibinfo{year}{1995}).

\bibitem[{\citenamefont{Steane}(1996)}]{Ste96}
\bibinfo{author}{\bibfnamefont{A.~M.} \bibnamefont{Steane}},
  \bibinfo{journal}{Phys.\ Rev.\ Lett.} \textbf{\bibinfo{volume}{{77}}},
  \bibinfo{pages}{793} (\bibinfo{year}{1996}).

\bibitem[{\citenamefont{Nielsen and Chuang}(2000)}]{NC00}
\bibinfo{author}{\bibfnamefont{M.~A.} \bibnamefont{Nielsen}} \bibnamefont{and}
  \bibinfo{author}{\bibfnamefont{I.~L.} \bibnamefont{Chuang}},
  \emph{\bibinfo{title}{Quantum Computation and Quantum Information}}
  (\bibinfo{publisher}{Cambridge University Press},
  \bibinfo{address}{Cambridge}, \bibinfo{year}{2000}).

\bibitem[{\citenamefont{Ozawa}(2003{\natexlab{a}})}]{03QLM}
\bibinfo{author}{\bibfnamefont{M.}~\bibnamefont{Ozawa}}, in
  \emph{\bibinfo{booktitle}{Proceedings of the Sixth International Conference
  on Quantum Communication, Measurement and Computing}}, edited by
  \bibinfo{editor}{\bibfnamefont{J.~H.} \bibnamefont{Shappiro}}
  \bibnamefont{and} \bibinfo{editor}{\bibfnamefont{O.}~\bibnamefont{Hirota}}
  (\bibinfo{publisher}{Rinton Press}, \bibinfo{address}{Princeton},
  \bibinfo{year}{2003}{\natexlab{a}}), pp. \bibinfo{pages}{175--180}.

\bibitem[{\citenamefont{Barnes and Warren}(1999)}]{BW99}
\bibinfo{author}{\bibfnamefont{J.~P.} \bibnamefont{Barnes}} \bibnamefont{and}
  \bibinfo{author}{\bibfnamefont{W.~S.} \bibnamefont{Warren}},
  \bibinfo{journal}{Phys.\ Rev.\ A} \textbf{\bibinfo{volume}{{60}}},
  \bibinfo{pages}{4363} (\bibinfo{year}{1999}).

\bibitem[{\citenamefont{Gea-Banacloche}(2002)}]{Ban02}
\bibinfo{author}{\bibfnamefont{J.}~\bibnamefont{Gea-Banacloche}},
  \bibinfo{journal}{Phys.\ Rev.\ A} \textbf{\bibinfo{volume}{{65}}},
  \bibinfo{pages}{022308} (\bibinfo{year}{2002}).

\bibitem[{\citenamefont{van Enk and Kimble}(2002)}]{EK02}
\bibinfo{author}{\bibfnamefont{S.~J.} \bibnamefont{van Enk}} \bibnamefont{and}
  \bibinfo{author}{\bibfnamefont{H.~J.} \bibnamefont{Kimble}},
  \bibinfo{journal}{Quantum Inf.\ Comput.} \textbf{\bibinfo{volume}{{2}}},
  \bibinfo{pages}{1} (\bibinfo{year}{2002}).

\bibitem[{\citenamefont{Ozawa}(2002{\natexlab{a}})}]{02CQC}
\bibinfo{author}{\bibfnamefont{M.}~\bibnamefont{Ozawa}},
  \bibinfo{journal}{Phys.\ Rev.\ Lett.} \textbf{\bibinfo{volume}{{\bf 89}}},
  \bibinfo{pages}{057902} (\bibinfo{year}{2002}{\natexlab{a}}).

\bibitem[{\citenamefont{Wigner}(1952)}]{Wig52}
\bibinfo{author}{\bibfnamefont{E.~P.} \bibnamefont{Wigner}},
  \bibinfo{journal}{Z. Phys.} \textbf{\bibinfo{volume}{{133}}},
  \bibinfo{pages}{101} (\bibinfo{year}{1952}).

\bibitem[{\citenamefont{Araki and Yanase}(1960)}]{AY60}
\bibinfo{author}{\bibfnamefont{H.}~\bibnamefont{Araki}} \bibnamefont{and}
  \bibinfo{author}{\bibfnamefont{M.~M.} \bibnamefont{Yanase}},
  \bibinfo{journal}{Phys.\ Rev.} \textbf{\bibinfo{volume}{{120}}},
  \bibinfo{pages}{622} (\bibinfo{year}{1960}).

\bibitem[{\citenamefont{Ozawa}(2003{\natexlab{b}})}]{03CQC(R)}
\bibinfo{author}{\bibfnamefont{M.}~\bibnamefont{Ozawa}},
  \bibinfo{journal}{Phys.\ Rev.\ Lett.} \textbf{\bibinfo{volume}{{\bf 91}}},
  \bibinfo{pages}{089802} (\bibinfo{year}{2003}{\natexlab{b}}).

\bibitem[{\citenamefont{Lidar}(2003)}]{Lid03}
\bibinfo{author}{\bibfnamefont{D.~A.} \bibnamefont{Lidar}},
  \bibinfo{journal}{Phys.\ Rev.\ Lett.} \textbf{\bibinfo{volume}{91}},
  \bibinfo{pages}{089801} (\bibinfo{year}{2003}).

\bibitem[{\citenamefont{Kawano and Ozawa}(2006)}]{06QGG}
\bibinfo{author}{\bibfnamefont{Y.}~\bibnamefont{Kawano}} \bibnamefont{and}
  \bibinfo{author}{\bibfnamefont{M.}~\bibnamefont{Ozawa}},
  \bibinfo{journal}{Phys.\ Rev.\ A} \textbf{\bibinfo{volume}{73}},
  \bibinfo{pages}{012339} (\bibinfo{year}{2006}).

\bibitem[{\citenamefont{Ozawa}(2002{\natexlab{b}})}]{02CLU}
\bibinfo{author}{\bibfnamefont{M.}~\bibnamefont{Ozawa}},
  \bibinfo{journal}{Phys.\ Rev.\ Lett.} \textbf{\bibinfo{volume}{{\bf 88}}},
  \bibinfo{pages}{050402} (\bibinfo{year}{2002}{\natexlab{b}}).

\bibitem[{\citenamefont{Ozawa}(2003{\natexlab{c}})}]{03UPQ}
\bibinfo{author}{\bibfnamefont{M.}~\bibnamefont{Ozawa}},
  \bibinfo{journal}{Int.\ J. Quant.\ Inf.} \textbf{\bibinfo{volume}{1}},
  \bibinfo{pages}{569} (\bibinfo{year}{2003}{\natexlab{c}}).

\bibitem[{\citenamefont{Ozawa}(2003{\natexlab{d}})}]{03UVR}
\bibinfo{author}{\bibfnamefont{M.}~\bibnamefont{Ozawa}},
  \bibinfo{journal}{Phys.\ Rev.\ A} \textbf{\bibinfo{volume}{67}},
  \bibinfo{pages}{042105} (\bibinfo{year}{2003}{\natexlab{d}}).

\bibitem[{\citenamefont{Ozawa}(2003{\natexlab{e}})}]{03HUR}
\bibinfo{author}{\bibfnamefont{M.}~\bibnamefont{Ozawa}},
  \bibinfo{journal}{Phys.\ Lett.\ A} \textbf{\bibinfo{volume}{318}},
  \bibinfo{pages}{21} (\bibinfo{year}{2003}{\natexlab{e}}).

\bibitem[{\citenamefont{Ozawa}(2004)}]{04URN}
\bibinfo{author}{\bibfnamefont{M.}~\bibnamefont{Ozawa}},
  \bibinfo{journal}{Ann.\ Phys.\ (N.Y.)} \textbf{\bibinfo{volume}{311}},
  \bibinfo{pages}{350} (\bibinfo{year}{2004}).

\bibitem[{\citenamefont{Gea-Banacloche and Ozawa}(2005)}]{05CQL}
\bibinfo{author}{\bibfnamefont{J.}~\bibnamefont{Gea-Banacloche}}
  \bibnamefont{and} \bibinfo{author}{\bibfnamefont{M.}~\bibnamefont{Ozawa}},
  \bibinfo{journal}{J. Opt.\ B: Quantum Semiclass.\ Opt.}
  \textbf{\bibinfo{volume}{7}}, \bibinfo{pages}{S326} (\bibinfo{year}{2005}).

\bibitem[{\citenamefont{Itano}(2003)}]{Ita03}
\bibinfo{author}{\bibfnamefont{W.~M.} \bibnamefont{Itano}},
  \bibinfo{journal}{Phys.\ Rev.\ A} \textbf{\bibinfo{volume}{{68}}},
  \bibinfo{pages}{046301} (\bibinfo{year}{2003}).

\bibitem[{\citenamefont{Silberfarb and Deutsch}(2004)}]{SD04}
\bibinfo{author}{\bibfnamefont{A.}~\bibnamefont{Silberfarb}} \bibnamefont{and}
  \bibinfo{author}{\bibfnamefont{I.~H.} \bibnamefont{Deutsch}},
  \bibinfo{journal}{Phys.\ Rev.\ A} \textbf{\bibinfo{volume}{69}},
  \bibinfo{pages}{042308} (\bibinfo{year}{2004}).

\bibitem[{\citenamefont{van Enk and Kimble}(2003)}]{EK03}
\bibinfo{author}{\bibfnamefont{S.~J.} \bibnamefont{van Enk}} \bibnamefont{and}
  \bibinfo{author}{\bibfnamefont{H.~J.} \bibnamefont{Kimble}},
  \bibinfo{journal}{Phys.\ Rev.\ A} \textbf{\bibinfo{volume}{{68}}},
  \bibinfo{pages}{046302} (\bibinfo{year}{2003}).

\bibitem[{\citenamefont{Gea-Banacloche}(2003)}]{Ban03}
\bibinfo{author}{\bibfnamefont{J.}~\bibnamefont{Gea-Banacloche}},
  \bibinfo{journal}{Phys.\ Rev.\ A} \textbf{\bibinfo{volume}{{68}}},
  \bibinfo{pages}{046303} (\bibinfo{year}{2003}).

\bibitem[{\citenamefont{Nielsen and Chuang}(1997)}]{NC97}
\bibinfo{author}{\bibfnamefont{M.~A.} \bibnamefont{Nielsen}} \bibnamefont{and}
  \bibinfo{author}{\bibfnamefont{I.~L.} \bibnamefont{Chuang}},
  \bibinfo{journal}{Phys.\ Rev.\ Lett.} \textbf{\bibinfo{volume}{79}},
  \bibinfo{pages}{321} (\bibinfo{year}{1997}).

\bibitem[{\citenamefont{Vidal and Cirac}(2000)}]{VC00}
\bibinfo{author}{\bibfnamefont{C.}~\bibnamefont{Vidal}} \bibnamefont{and}
  \bibinfo{author}{\bibfnamefont{J.~I.} \bibnamefont{Cirac}},
  \emph{\bibinfo{title}{Storage of quantum dynamics in quantum states: a
  quasi-perfect programmable quantum gate}} (\bibinfo{year}{2000}),
  \bibinfo{note}{e-print quant-ph/0012067}.

\bibitem[{\citenamefont{Hillery et~al.}(2006)\citenamefont{Hillery, Ziman, and
  Bu\v{z}ek}}]{HZB06}
\bibinfo{author}{\bibfnamefont{M.}~\bibnamefont{Hillery}},
  \bibinfo{author}{\bibfnamefont{M.}~\bibnamefont{Ziman}}, \bibnamefont{and}
  \bibinfo{author}{\bibfnamefont{V.}~\bibnamefont{Bu\v{z}ek}},
  \bibinfo{journal}{Phys.\ Rev.\ A} \textbf{\bibinfo{volume}{73}},
  \bibinfo{pages}{022345} (\bibinfo{year}{2006}).

\bibitem[{\citenamefont{D'Ariano and Perinotti}(2005{\natexlab{a}})}]{DP05}
\bibinfo{author}{\bibfnamefont{G.~M.} \bibnamefont{D'Ariano}} \bibnamefont{and}
  \bibinfo{author}{\bibfnamefont{P.}~\bibnamefont{Perinotti}},
  \bibinfo{journal}{Phys.\ Rev.\ Lett.} \textbf{\bibinfo{volume}{94}},
  \bibinfo{pages}{090401} (\bibinfo{year}{2005}{\natexlab{a}}).

\bibitem[{\citenamefont{D'Ariano and Perinotti}(2005{\natexlab{b}})}]{DP05b}
\bibinfo{author}{\bibfnamefont{G.~M.} \bibnamefont{D'Ariano}} \bibnamefont{and}
  \bibinfo{author}{\bibfnamefont{P.}~\bibnamefont{Perinotti}},
  \emph{\bibinfo{title}{On the most efficient unitary transformation for
  programming quantum channels}} (\bibinfo{year}{2005}{\natexlab{b}}),
  \bibinfo{note}{e-print quant-ph/0509183}.

\bibitem[{\citenamefont{D'Ariano and Perinotti}(2005{\natexlab{c}})}]{DP05c}
\bibinfo{author}{\bibfnamefont{G.~M.} \bibnamefont{D'Ariano}} \bibnamefont{and}
  \bibinfo{author}{\bibfnamefont{P.}~\bibnamefont{Perinotti}},
  \emph{\bibinfo{title}{Programmable quantum channels and measurements}}
  (\bibinfo{year}{2005}{\natexlab{c}}), \bibinfo{note}{e-print
  quant-ph/0510033}.

\bibitem[{\citenamefont{Paulsen}(1986)}]{Pau86}
\bibinfo{author}{\bibfnamefont{V.~I.} \bibnamefont{Paulsen}},
  \emph{\bibinfo{title}{Completely bounded maps and dilations}}, Pitman Resarch
  Notes in Math. {146} (\bibinfo{publisher}{Longman}, \bibinfo{address}{New
  York}, \bibinfo{year}{1986}).

\bibitem[{\citenamefont{Belavkin et~al.}(2005)\citenamefont{Belavkin, D'Ariano,
  and Raginsky}}]{BDR05}
\bibinfo{author}{\bibfnamefont{V.~P.} \bibnamefont{Belavkin}},
  \bibinfo{author}{\bibfnamefont{G.~M.} \bibnamefont{D'Ariano}},
  \bibnamefont{and} \bibinfo{author}{\bibfnamefont{M.}~\bibnamefont{Raginsky}},
  \bibinfo{journal}{J. Math.\ Phys.} \textbf{\bibinfo{volume}{46}},
  \bibinfo{pages}{062106} (\bibinfo{year}{2005}).

\bibitem[{\citenamefont{Hotta et~al.}(2005)\citenamefont{Hotta, Karasawa, and
  Ozawa}}]{05AAE}
\bibinfo{author}{\bibfnamefont{M.}~\bibnamefont{Hotta}},
  \bibinfo{author}{\bibfnamefont{T.}~\bibnamefont{Karasawa}}, \bibnamefont{and}
  \bibinfo{author}{\bibfnamefont{M.}~\bibnamefont{Ozawa}},
  \bibinfo{journal}{Phys.\ Rev.\ A} \textbf{\bibinfo{volume}{{\bf 72}}},
  \bibinfo{pages}{052334} (\bibinfo{year}{2005}).

\bibitem[{\citenamefont{Szego}(1967)}]{Sze67}
\bibinfo{author}{\bibfnamefont{G.}~\bibnamefont{Szego}},
  \emph{\bibinfo{title}{Orthogonal Polynomials}} (\bibinfo{publisher}{American
  Mathematical Society}, \bibinfo{address}{Providence, R.I.},
  \bibinfo{year}{1967}).

\bibitem[{\citenamefont{Chihara}(1978)}]{Chi78}
\bibinfo{author}{\bibfnamefont{T.~S.} \bibnamefont{Chihara}},
  \emph{\bibinfo{title}{An Introduction to Orthogonal Polynomials}}
  (\bibinfo{publisher}{Gordon and Breach}, \bibinfo{address}{New York},
  \bibinfo{year}{1978}).

\end{thebibliography}
\end{document}